\documentclass[12pt]{article}

\usepackage{amsmath,amssymb,epsfig,bbm,color}
\usepackage{epsfig}
\usepackage{euscript}
\usepackage{cite}

\newcommand{\Kcal}{{\cal K}}

\newcommand{\Peu}{\EuScript{P}}
\newcommand{\Keu}{\EuScript{K}}

\newcommand{\veps}{\varepsilon}

\newcommand{\tB}{{\bar{t}}}

\def\beeq{\begin{eqnarray}}
\def\eeeq{\end{eqnarray}}
\def\be{\begin{equation}}
\def\ee{\end{equation}}

\def\fprime{f^\prime}
\def\eto{{\rm{e}}}

\begin{document}

%%%%%%%%%%%%%%%%%%%%%%%%%%%%%%%%%%%%%%%%%%%%%%%%%%%%%%%%%%%%%%%%%%%%%%%%%%%%%%
%%%%%%%%%%%%%%%%%%%%%%%%%%%%%%%%%%%%%%%%%%%%%%%%%%%%%%%%%%%%%%%%%%%%%%%%%%%%%%%
\begin{titlepage}

\begin{flushright}
{\bf IFJPAN-IV-07-09\\
    CERN-PH-TH/2007-142}
\end{flushright}

\vspace{1mm}
\begin{center}
  {\LARGE\bf%
    Solving QCD evolution equations\\
    in rapidity space with\strut \\
    Markovian Monte Carlo%
    $^{\star}$
}
\end{center}
\vspace{2mm}

\begin{center}
{\large\bf  K.~Golec-Biernat$^{ac}$,S.~Jadach$^{ad}$,\\
  W.~P\l{}aczek$^b$ {\rm and} M.~Skrzypek$^{ad}$}
\end{center}

\vspace{2mm}
\begin{center}
{\em $^a$Institute of Nuclear Physics, Polish Academy of Sciences,\\
  ul.\ Radzikowskiego 152, 31-342 Cracow, Poland.}\\ \vspace{2mm}
{\em $^b$Marian Smoluchowski Institute of Physics, Jagiellonian University,\\
   ul.\ Reymonta 4, 30-059 Cracow, Poland.}\\ \vspace{2mm}
{\em $^c$Institute of Physics, University of Rzeszow,\\
   ul.\ Rejtana 16A, 35-959 Rzeszow, Poland.}
\\ \vspace{2mm}
{\em $^d$CERN, PH Department, CH-1211 Geneva 23, Switzerland.} 
\end{center}

\vspace{5mm}
\begin{abstract}
This work covers methodology of solving QCD evolution equation
of the parton distribution 
using Markovian Monte Carlo (MMC) algorithms
in a class of models ranging from DGLAP to CCFM.
One of the purposes of the above MMCs is to test 
the other more sophisticated Monte Carlo programs, 
the so-called Constrained Monte Carlo (CMC) programs,
which will be used as a building block in the parton shower MC.
This is why the mapping of the evolution variables 
(eikonal variable and evolution time)
into four-momenta is also defined and tested.
The evolution time is identified with the rapidity variable
of the emitted parton.
The presented MMCs are tested independently,
with $\sim 0.1\%$ precision,
against the non-MC program {\tt APCheb}
especially devised for this purpose.
\end{abstract}

\vspace{1mm}
\begin{center}
\em Submitted to Acta Physica Polonica
\end{center}

\vspace{5mm}
\begin{flushleft}
{\bf IFJPAN-IV-07-09\\
     CERN-PH-TH/2007-142\\
     August~2007}
\end{flushleft}

\vspace{3mm}
\footnoterule
\noindent
{\footnotesize
$^{\star}$This work is partly supported by the EU grant MTKD-CT-2004-510126
 in partnership with the CERN Physics Department and by the Polish Ministry
 of Scientific Research and Information Technology grant No 620/E-77/6.PR
 UE/DIE 188/2005-2008.
}

\end{titlepage}
%%%%%%%%%%%%%%%%%%%%%%%%%%%%%%%%%%%%%%%%%%%%%%%%%%%%%%%%%%%%%%%%%%%%%%%%%%%%%%%
%%%%%%%%%%%%%%%%%%%%%%%%%%%%%%%%%%%%%%%%%%%%%%%%%%%%%%%%%%%%%%%%%%%%%%%%%%%%%%%

\section{Introduction}

The problem of solving numerically the so-called evolution
equations of the parton distribution functions (PDFs) in
quantum Chromodynamics (QCD) is revisited again and again
in all effort of providing more precise perturbative QCD predictions
for the experiments in the Large Hadron Collider (LHC)
and other hadron colliders (e.g.\ Tevatron).
In this work we intend to present a methodology of solving
QCD evolution equations using Monte Carlo techniques for
several types of the evolutions, the resulting numerical results,
including the comparisons with other non-MC numerical methods.

Two decades ago, when first attempts of solving numerically
and precisely the evolution time dependence
of the parton distribution functions (PDFs)
according to the DGLAP~\cite{DGLAP} equations were made, 
it was unthinkable
that the Monte Carlo techniques could be used for this purpose.
It was simply because the
computers were too slow by several orders of the magnitude.
Instead, various faster techniques were developed, 
based mainly on dividing the evolution
time into short periods and using discrete grid in $x$-space --
they are presently still widely used.
Nowdays, with much faster computers, it is perfectly feasible 
to solving numerically the QCD evolution equations with 3--4 digit precision
for DGLAP and other types of evolutions,
albeit it is still much slower than with other techniques.

One may therefore ask the following question:
does the MC technique of solving QCD evolution equations have some advantages
over other techniques which makes it worth to pursue in spite of its slowness?
In our opinion the MC technique offers certain unique advantages.
Let us mention the most important ones:
Although numerical statistical error is usually bigger than for other methods,
this error is very stable and robust, not prone to any effects related
to finite grid or time slicing.
Another advantage of the MC method is that for many types of partons
one may solve the evolution equations for all parton types simultaneously,
without the need of diagonalizing kernels, 
that is using PDFs in the basis of gluon,
singlet quark and several types of the non-singlet quark components,
and then recombining that back.
Finally, the biggest potential advantage is that in the MC method
one can devise mapping of the evolution time and other variables
into four-momenta, hence to set-up the starting point for
constructing a more realistic treatment of the multiparton emission shower,
thet is the so-called parton shower MC.
Also, the extensions from orthodox DGLAP towards more complicated
kernels/evolutions featuring small $x$ resummations, such as CCFM~\cite{CCFM},
can be treated with the MC techniques more easily than with other methods.

It should be stressed that this work is closely related with another
work of ref.~\cite{Jadach:2007singleCMC}. In fact the MC programs of this
work are exploited in ref.~\cite{Jadach:2007singleCMC} to test more complicated
MC techniques of solving evolution equations.
The main difference between this work and ref.~\cite{Jadach:2007singleCMC}
is that here we concentrate on the Markovian class of MC solutions,
while ref.~\cite{Jadach:2007singleCMC} elaborates on the class
of non-Markovian techniques,
in which the parton energy fraction $x$ and its type $f$ are constrained (predefined).
The Markovian MC is better suited for the final-state parton
cascade while the constrained MC of ref.~\cite{Jadach:2007singleCMC} is better
for the initial state cascade,
for instance in hadron colliders ($W$/$Z$ boson production).

Our paper is organized as follows: 
In Section~2 we present general form
of evolution equations and their iterative solutions. 
In Section~3 we describe in detail three Markovian algorithms for solving
these equations. Section~4 contains details on evolution kernels 
and form-factors. In Section~5 we give some remarks on Monte Carlo implementations
of the above algorithms. Section~6 is devoted to the Chebyshev polynomials method 
of solving the evolution equations. 
In Section~7 we present our numerical results. Finally, Section~8 summerizes
the paper.

%%%%%%%%%%%%%%%%%%%%%%%%%%%%%%%%%%%%%%%%%%%%%%%%%%%%%%%%%%%%%%%%%%%%%%%%%%%%%%%
\section{General evolution equations}
In this work we shall cover several types of the QCD evolution equations
ranging from DGLAP~\cite{DGLAP} to CCFM~\cite{CCFM} and their extensions.
The generic evolution equation covering all types of QCD evolution
of our interest reads
\begin{equation}
\label{eq:genevoleq}
  \partial_t D_f(t,x)
 = \sum_{f'} \int_x^1 du\; \Keu_{ff'}(t,x,u) D_{f'}(t,u).
\end{equation}
The parton distribution function (PDF) is $D_j(t,u)$,
with $x$ being the fraction of the hadron momentum%
\footnote{Or, equivalently, the fraction of the eikonal ``plus'' variable.}
carried by the parton
and $j$ being the type (flavour) of the parton.
The so called evolution time $t=\ln Q$ represents in QCD logarithm
of the energy scale $Q=\mu$ determined by hard scattering process probing PDF.
%%%
The case of the LL DGLAP case \cite{DGLAP} is recovered
with the following identification
\begin{equation}
\label{eq:dglap-kernel}
  \Keu_{ff'}(t,x,u) = \frac{1}{u}{\Peu}_{ff'}\left(t,\frac{x}{u}\right)
  = \frac{\alpha_S(t)}{2\pi} \frac{2}{u} P^{(0)}_{ff'}\left(t,\frac{x}{u}\right),
\end{equation}
where $P^{(0)}_{ff'}(z)$ is the lowest order DGLAP kernel.

In the compact operator (matrix) notation eq.~(\ref{eq:genevoleq}) reads
\begin{equation}
\label{eq:op-evoleq}
  \partial_t {\bf D}(t) = {\bf K}(t) \; {\bf D}(t).
\end{equation}
%%%
Given a known ${\bf D}(t_0)$ at the initial time $t_0$, 
the formal solution at any later time $t\geq t_0$
is provided by the time ordered exponential
\begin{equation}
\label{eq:timex} 
  {\bf D}(t) 
   = \exp\left( \int_{t_0}^t {\bf K}(t') dt' \right)_{T.O.} \!\! {\bf D}(t_0)
   = {\bf G}_{\bf K}(t,t_0) {\bf D}(t_0).
\end{equation}
The {\em time-ordered exponential} evolution operator reads%
\footnote{Here and in the following we adopt the following conventions
 $\prod_{i=1}^n A_i \equiv A_n A_{n-1}\dots A_2A_1$ and
 $\prod_{i=1}^n \int dt_i 
  \equiv \int dt_n \int dt_{n-1}\dots \int dt_2 \int dt_1$.
 The inverse ordering will be similarly denoted
 with $\overline{\prod}_{i=1}^{n}$.}
\begin{equation}
\label{eq:torder}
{\bf G}_{\bf K}(t,t_0)=
{\bf G}({\bf K};t,t_0)=
\exp\left( \int_{t_0}^t {\bf K}(t') dt' \right)_{T.O.}
 = {\bf I} +\sum_{n=1}^\infty 
      \prod_{i=1}^n  \int_{t_0}^t dt_i \theta_{t_i>t_{i-1}} {\bf K}(t_i),
\end{equation}
where $ ({\bf I})_{f_2,f_1}(x_2,x_1)\equiv \delta_{f_2f_1}\delta_{x_2=x_1}$ and
the multiplication of the operators is defined as follows
\begin{equation}
\label{eq:clarity}
   \big( {\bf K}(t_2){\bf K}(t_1)\big)_{f_2,f_1}(x_2,x_1)=
   \sum_{f'} \int_{x_2}^{x_1} dx'\;
        \Keu_{f_2f'}(t_2,x_2,x') \Keu_{f'f_1}(t_1,x',x_1).
\end{equation}
From now on we adopt the following notation%
\footnote{Similarly, we define $\theta_{z<y<x}=\theta_{z<y}\theta_{y<x}$.}:
\[
  \delta_{x=y}=\delta(x-y),\qquad
  \theta_{y<x}=1~~~\hbox{\rm for}~~~y<x~~~\hbox{\rm and}~~~
  \theta_{y<x}=0~~~\hbox{\rm for}~~~y\geq x.
\]

In the case of the kernel split into two components,
$ {\bf K}(t)={\bf K}^A(t)+{\bf K}^B(t), $
the solution of eq.~(\ref{eq:timex})
can be reorganized as follows%
\footnote{ The scope of the index $i$ in $\prod_i$ ceases at
  the closing bracket, but validity scope of indiced variables,
  like $t_i$, extends until the formula's end.
  The use of eq.~(\ref{eq:clarity}) is understood accordingly.
}
\begin{equation}
\label{eq:timex2}
\begin{split}
&{\bf D}(t)
={\bf G}_{{\bf K}^B}(t,t_0)\; {\bf D}(t_0)
+\sum_{n=1}^\infty
 \left[\prod_{i=1}^n \int_{t_{i-1}}^{t}dt_i\right]
 {\bf G}_{{\bf K}^B}(t,t_{n})\; 
 \left[\prod_{i=1}^n
    {\bf K}^A(t_{i})
    {\bf G}_{{\bf K}^B}(t_{i},t_{i-1})
 \right]
{\bf D}(t_0),
\\&
{\bf G}_{{\bf K}}(t,t_{0})
={\bf G}_{{\bf K}^B}(t,t_0)\;
+\sum_{n=1}^\infty
 \left[\prod_{i=1}^n \int_{t_{i-1}}^{t}dt_i\right]
 {\bf G}_{{\bf K}^B}(t,t_{n})\; 
 \left[\prod_{i=1}^n
    {\bf K}^A(t_{i})
    {\bf G}_{{\bf K}^B}(t_{i},t_{i-1})
 \right],
\end{split}
\end{equation}
where ${\bf G}_{{\bf K}^B}$ is the evolution operator
of eq.~(\ref{eq:torder}) of the evolution with the kernel ${\bf K}^B$.
Formal proof of identities in
eq.~(\ref{eq:timex2}) can be found in  ref.~\cite{Jadach:2006yr}.

%%%%%%%%%%%%%%%%%%%%%%%%%%%%%%%%%%%%%%%%%%%%%
\subsection{Resuming virtual corrections}
Monte Carlo method cannot efficiently deal with the non-positive distributions,
hence resummation of negative virtual part in the evolution
kernel is a necessary preparatory step.
It will be done with help of identity of eq.~(\ref{eq:timex2}).
We are going resum (negative) diagonal virtual part
${\bf K}^V={\bf K}^B$ in the kernel
\begin{equation}
\label{eq:evkernel}
\begin{split}
&\Keu_{ff'}(t,x,u)= \Keu^V_{ff'}(t,x,u)+\Keu^R_{ff'}(t,x,u),\quad
 \Keu^V_{ff'}(t,x,u)= -\delta_{ff'}\delta_{x=u}\Keu^v_{ff}(t,x).
\end{split}
\end{equation}
At this point we do not need to be very specific about $\Keu^R_{ff'}(t,x,u)$
-- we only remark that due to infrared (IR) singularity
at $x=u$ and $f=f'$ it includes IR cut-off, typically $u-x>\Delta(x,u,t)$,
causing $\Keu^v$ to be also $\Delta$-dependent.

Thanks to diagonality of the kernel ${\bf K}^V$,
the corresponding time-ordered exponential is easily calculable
\begin{equation}
\label{eq:phidef}
\{{\bf G}_{{\bf K}^V}(t,t')\}_{ff'}(x,u)
  =\delta_{ff'}\delta_{x=u}\; e^{-\Phi_f(t,t'|x)},\qquad
\Phi_f(t,t'|x)=\int_{t'}^{t} dt''\; \Keu^v_{ff}(t'',x).
\end{equation}
Inserting the above in eq.~(\ref{eq:timex2}) we obtain
\begin{equation}
\label{eq:timex3}
{\bf D}(t)=
\sum_{n=0}^\infty
 \left[\prod_{i=1}^n \int_{t_{i-1}}^{t}dt_i\right]
 {\bf G}_{{\bf K}^V}(t,t_{n})\; 
 \left[\prod_{i=1}^n
    {\bf K}^R(t_{i})
    {\bf G}_{{\bf K}^V}(t_{i},t_{i-1})
 \right]
{\bf D}(t_0).
\end{equation}
More compact notation is obtained with the
prescription $\prod_{i=k}^{k-1} {\bf A}_i\equiv {\bf I}$
and $\prod_{i=k}^{k-1}\int dt_i\equiv 1$.

%%%%%%%%%%%%%%%%%%%%%%%%%%%%%%%%%%%%%%%%%%%%%
\subsection{Momentum sum rule}
Evolution equations and their time ordered solutions
do not require any assumptions about the normalization of PDFs
and kernels.
However, Markovian Monte Carlo methods are inherently based
on the unitary normalization of the probability distributions
(for the forward step).
Hence, we concentrate on the evolution equations
which are supplemented with some conservation rule,
providing time-independent normalization condition.
For DGLAP it is the momentum sum rule
which is obeyed exactly and is exploited to this end
(it can also be used for the CCFM class models).
It will be also formulated in terms of the compact operator formalism.
Let us define operator (vector) $\bar{\bf E}$ acting from the left side
\begin{equation}
\label{eq:momsumrule}
  \bar{\bf E}\; {\bf D}(t)
  \equiv \int_0^1 dx \sum_{f} x\; D_f(t,x).
\end{equation}
The momentum sum rule can be stated as the following time conservation law:
\begin{equation}
  \partial_t \bar{\bf E}\; {\bf D}(t) =0.
\end{equation}
Inserting evolution equation one obtains immediately
\begin{equation}
  \partial_t \bar{\bf E}\; {\bf D}(t)
= \bar{\bf E} {\bf K}\; {\bf D}(t)=0.
\end{equation}
The sufficient condition for the above to be true
is the following property of the kernel
\begin{equation}
\label{eq:kernelsum}
 \bar{\bf E} {\bf K}=\bar{\bf 0},\quad
 (\bar{\bf E} {\bf K})_f(u)=
 \sum_{f'}\int_0^1 dx\; x\Keu_{f'f}(t,x,u)  =0,
\end{equation}
for any $u$ and $f$. In particular we have
$\bar{\bf E} {\bf K^V}+\bar{\bf E} {\bf K^R}=\bar{\bf 0}$,
from which we can derive immediately the virtual part of the kernel
\begin{equation}
\label{eq:kernv}
 -(\bar{\bf E} {\bf K^V})_f(u)
 =u\Keu^v_{ff}(t,u) 
 = \sum_{f'} \int^u_0 
  d x\; x \; \Keu^{R}_{f'f} (t, x, u)
 =(\bar{\bf E} {\bf K^R})_f(u).
\end{equation}
From $\bar{\bf E} {\bf K}=\bar{\bf 0}$
also follows the following usefull identity
\begin{equation}
 \bar{\bf E} {\bf G}_{\bf K}(t,t_0)=\bar{\bf E},
\end{equation}
which provides immediately
$\bar{\bf E} {\bf D}(t)=\bar{\bf E} {\bf D}(t_0)$.

%%%%%%%%%%%%%%%%%%%%%%%%%%%%%%%%%%%%%%%%%%%%%
\subsection{Markovianization}
The aim is now to transform eq.~(\ref{eq:timex3}) into
a form better suited for the Monte Carlo evaluation, using Markovian algorithm.
The basic problem is to show how to change the integration
order from
$\int^{t}_{t_{0}} dt_n \dots \int^{t_{3}}_{t_{0}} dt_2 \int^{t_{2}}_{t_{0}} dt_1$
to
$\int^{t}_{t_{0}} dt_1\int^{t_{1}}_{t_{0}} dt_2 \dots\int^{t_{n-1}}_{t_{0}} dt_n$,
taking into account non-commutative character of the product of the kernels
in the time ordered exponentials.

It is convenient not only to  change the order of the $t$-integration
but also to transpose simultaneously (temporarily) both sides of eq.~(\ref{eq:timex3})
\begin{equation}
\label{eq:timex4}
\begin{split}
&
\bar{\bf D}(t)=
\bar{\bf D}(t_0)
\sum_{n=0}^\infty
 \left[
 \overline{\prod}_{i=1}^{n}
    \int_{t_{0}}^{t}dt_i\; \theta_{t_i>t_{i-1}}\;
    \bar{\bf G}_{{\bf K}^V}(t_{i},t_{i-1})
    \bar{\bf K}^R(t_{i})
 \right]
 \bar{\bf G}_{{\bf K}^V}(t,t_{n}).
\end{split}
\end{equation}
In the next step we isolate the integration over $t_1$,
the outermost one,
\begin{equation}
\label{eq:timex5}
\begin{split}
&
\bar{\bf D}(t)=
\bar{\bf D}(t_0)
\bigg\{
 \bar{\bf G}_{{\bf K}^V}(t,t_{0})
+\int_{t_{0}}^{t}dt_1\;
 \bar{\bf G}_{{\bf K}^V}(t_{1},t_{0})
 \bar{\bf K}^R(t_{1})
\\&~~~~~~~~~~~~\times
 \sum_{n=1}^\infty
 \left[
 \overline{\prod}_{i=2}^{n}
    \int_{t_{1}}^{t}dt_i\; \theta_{t_i>t_{i-1}}\;
    \bar{\bf G}_{{\bf K}^V}(t_{i},t_{i-1})
    \bar{\bf K}^R(t_{i})
 \right]
 \bar{\bf G}_{{\bf K}^V}(t,t_{n})
\bigg\}.
\end{split}
\end{equation}
Closer look into second line in the above equation reveals%
\footnote{ After renaming $t_i\rightarrow t_{i-1}$ 
   and shifting indices $i$ and $n$ by one.}
that it represents again the time ordered evolution operator
$\bar{\bf G}_{\bf K}(t,t_1)$ (with $t_0\rightarrow t_1$).
We obtain therefore
\begin{equation}
\label{eq:timex6}
\begin{split}
& \bar{\bf D}(t)=
\bar{\bf D}(t_0)
\bigg\{
 \bar{\bf G}_{{\bf K}^V}(t,t_{0})
+\int_{t_{0}}^{t}dt_1\;
 \bar{\bf G}_{{\bf K}^V}(t_{1},t_{0})
 \bar{\bf K}^R(t_{1})\;\;
 \bar{\bf G}_{\bf K}(t,t_1)
\bigg\}.
\end{split}
\end{equation}
Transposition can be now removed and the integral over $t_1$ is pulled out
\begin{equation}
\label{eq:evol0}
\begin{split}
& {\bf D}(t)=
\int_{t_{0}}^{t}dt_1\;
\bigg\{
{\bf G}_{\bf K}(t,t_1)\;\;
 {\bf K}^R(t_{1})
 {\bf G}_{{\bf K}^V}(t_{1},t_{0})
+{\bf G}_{{\bf K}^V}(t,t_{0})\delta_{t_1=t}
\bigg\}
{\bf D}(t_0)
\end{split}
\end{equation}
The above result can be also presented
as an integral equation for the evolution operator
\begin{equation}
\label{eq:evol1}
\begin{split}
& {\bf G}_{{\bf K}}(t,t_0)=
\int_{t_{0}}^{t}dt_1\;
\bigg\{
 {\bf G}_{\bf K}(t,t_1)\;\;
 {\bf K}^R(t_{1})
 {\bf G}_{{\bf K}^V}(t_{1},t_{0})
+{\bf G}_{{\bf K}^V}(t,t_{0})\delta_{t_1=t}
\bigg\}
\end{split}
\end{equation}
This can be inserted back into eq.~(\ref{eq:timex6}) many times.
The following example shows three levels of the nesting
\begin{equation}
\label{eq:evol2}
\begin{split}
{\bf D}(t)=
\int_{t_{0}}^{t} dt_1 \bigg(
\int_{t_{1}}^{t} dt_2 \bigg[
\int_{t_{2}}^{t} dt_3 \bigg\{
   {\bf G}_{\bf K}(t,t_3)\;\;
&  {\bf K}^R(t_{3})
   {\bf G}_{{\bf K}^V}(t_{3},t_{2})
  +{\bf G}_{{\bf K}^V}(t,t_{2})\delta_{t_3=t}
\bigg\}
\\ \times
&  {\bf K}^R(t_{2})
   {\bf G}_{{\bf K}^V}(t_{2},t_{1})
  +{\bf G}_{{\bf K}^V}(t,t_{1})\delta_{t_2=t}
\bigg]
\\ \times
&
 {\bf K}^R(t_{1})
 {\bf G}_{{\bf K}^V}(t_{1},t_{0})
+{\bf G}_{{\bf K}^V}(t,t_{0})\delta_{t_1=t}
\bigg)
{\bf D}(t_0).
\end{split}
\end{equation}
It should be stressed that integration over $t_1$ is now the external
one and in the MC it will be generated as a first one.

If the above nesting is continued to the level $N+1$,
then one may argue that the contribution
from the term with ${\bf G}_{\bf K}(t,t_{N+1})$
for large $N$ decreases like $1/N!$, hence
in the Markovian MC we may use
the following formula ``truncated'' at large fixed $N$
playing a role of a dummy technical parameter:
\begin{equation}
\label{eq:evol3}
\begin{split}
{\bf D}(t)=
\int_{t_{0}}^{t} dt_1    \bigg(
\int_{t_{1}}^{t} dt_2    \bigg[
\int_{t_{2}}^{t} dt_3    \bigg\{& \dots
\\
\dots
\int_{t_{N-1}}^{t} dt_{N}\bigg\{
&  {\bf K}^R(t_{N})
   {\bf G}_{{\bf K}^V}(t_{N},t_{N-1})
  +{\bf G}_{{\bf K}^V}(t,t_{N-1})\delta_{t_{N}=t}
\bigg\}
\\~~~~~~~~~~~~~ \vdots
\\ \times
&  {\bf K}^R(t_{2})
   {\bf G}_{{\bf K}^V}(t_{2},t_{1})
  +{\bf G}_{{\bf K}^V}(t,t_{1})\delta_{t_2=t}
\bigg]
\\ \times
&
 {\bf K}^R(t_{1})
 {\bf G}_{{\bf K}^V}(t_{1},t_{0})
+{\bf G}_{{\bf K}^V}(t,t_{0})\delta_{t_1=t}
\bigg)
{\bf D}(t_0),
\end{split}
\end{equation}
where the integration over $t_{N+1}$ was consumed
by $\delta_{t_{N+1}=t}$.
The above identity will be instrumental in constructing MMC algorithm
in the following section.

%%%%%%%%%%%%%%%%%%%%%%%%%%%%%%%%%%%%%%%%%%%%%%%%%%%%%%%%%%%%%%%%%%%%%%%%%%%%%%%
\section{Markovian MC algorithms}
For the Monte Carlo method one needs
a (sum of) scalar multi-dimensional integral.
For the straightforward Markovian algorithm we
shall take the following multi-integral
\begin{equation}
C =\bar{\bf E} {\bf D}(t)
  = \bar{\bf E} {\bf G}_{\bf K}(t,t_0) {\bf D}(t_0).
\end{equation}
The aim is to generate with the MC method all internal
integration variables in the above equation.
Then, the histogram of the variable $x=x_n$ and
flavour type $f=f_n$ is evaluated in the high statistic MC run.
Such a histogram is defined by means of inserting Dirac delta
functions in the above multi-integral:
\begin{equation}
D_f(x)=
  \sum_{n=0}^\infty
  \sum_{f_n f_0} \int dx_n dx_0\;
  \Big({\bf G}_{\bf K}(t,t_0)\Big)^{(n)}_{f_n,f_0}(x_n,x_0)\;
  \delta_{x=x_n}\delta_{f f_n}\;
  D_{f_0}(t_0,x_0),
\end{equation}
where $n$ is the dimensionality of the integral in ${\bf G}_{\bf K}$.

%%%%%%%%%%%%%%%%%%%%%%%%%%%%%%%%%%%%%%%%%%
\subsection{Basic formalism}
As a warm-up exercise let us insert ${\bf D}(t)$
of eq.~(\ref{eq:evol0}) into $\bar{\bf E} {\bf D}(t)$
and check how the
identity $ \bar{\bf E} {\bf D}(t)=\bar{\bf E}{\bf D}(t_0)$ is recovered
through explicit integration over $t_1$
\begin{equation}
\label{eq:evol4}
\begin{split}
\bar{\bf E}{\bf D}(t)
&=\int_{t_{0}}^{t}dt_1\;
\bigg\{
 \bar{\bf E}{\bf G}_{\bf K}(t,t_1)\;\;
 {\bf K}^R(t_{1})
 {\bf G}_{{\bf K}^V}(t_{1},t_{0})
+\bar{\bf E}{\bf G}_{{\bf K}^V}(t,t_{0})\delta_{t_1=t}
\bigg\}
{\bf D}(t_0)
\\&
=\int_{t_{0}}^{t}dt_1\;
\bigg\{
-\bar{\bf E}\;{\bf K}^V(t_{1})
 {\bf G}_{{\bf K}^V}(t_{1},t_{0})
+\bar{\bf E}{\bf G}_{{\bf K}^V}(t,t_{0})\delta_{t_1=t}
\bigg\}
{\bf D}(t_0)
\\&
=\int_{t_{0}}^{t}dt_1\;
\bigg\{
-\bar{\bf E}\;\partial_{t_1}
 {\bf G}_{{\bf K}^V}(t_{1},t_{0})
+\bar{\bf E}{\bf G}_{{\bf K}^V}(t,t_{0})\delta_{t_1=t}
\bigg\}
{\bf D}(t_0)
\\&
=\bigg\{
-\bar{\bf E}{\bf G}_{{\bf K}^V}(t_{1},t_{0})|_{t_1=t_0}^{t_1=t}
+\bar{\bf E}{\bf G}_{{\bf K}^V}(t,t_{0})
\bigg\}
{\bf D}(t_0)
=\bar{\bf E}\;{\bf D}(t_0).
\end{split}
\end{equation}
In the above the most essential was
the use of $\bar{\bf E}{\bf G}_{\bf K}(t,t_1)=\bar{\bf E}$
in the first step,
because it has allowed to decouple $t_1$-integration from
the integrations inside ${\bf G}_{\bf K}(t,t_1)$.
Next,  $\bar{\bf E}{\bf K}^R(t_{1})=-\bar{\bf E}{\bf K}^V(t_{1})$ was employed,
then the evolution equation for $ {\bf G}_{{\bf K}^V}$
and finally ${\bf G}_{{\bf K}^V}(t_{0},t_{0})={\bf I}$ was also used.
The decoupled inner integrations are explicitly present in the
following iterative formula
\begin{equation}
\label{eq:evol5}
\begin{split}
\bar{\bf E}{\bf D}(t)=
\int_{t_{0}}^{t} dt_1    \bigg(
\int_{t_{1}}^{t} dt_2    \bigg[&
\int_{t_{2}}^{t} dt_3    \bigg\{ \dots
\\
\dots
\int_{t_{N-1}}^{t} dt_{N}\bigg\{
&  \bar{\bf E}{\bf K}^R(t_{N})
   {\bf G}_{{\bf K}^V}(t_{N},t_{N-1})
  +\bar{\bf E}{\bf G}_{{\bf K}^V}(t,t_{N-1})\delta_{t_{N}=t}
\bigg\}
\\~~~~~~~~~~~~~ \vdots
\\ \times
&  {\bf K}^R(t_{2})
   {\bf G}_{{\bf K}^V}(t_{2},t_{1})
  +\bar{\bf E}{\bf G}_{{\bf K}^V}(t,t_{1})\delta_{t_2=t}
\bigg]
\\ \times
&
 {\bf K}^R(t_{1})
 {\bf G}_{{\bf K}^V}(t_{1},t_{0})
+\bar{\bf E}{\bf G}_{{\bf K}^V}(t,t_{0})\delta_{t_1=t}
\bigg)
{\bf D}(t_0).
\end{split}
\end{equation}
Again, we would like to stress that the order of the integration
starting from $t_1$ and ending with $t_N$ is exactly the one which will be
realized in the Markovian Monte Carlo algorithm.

\subsection{Straightforward Markovian algorithm}
%%%%%%%%%%%%%%%%%%%%%%%%%%%%%%%%%%%%%%%%%%%%%%%%%%%%%%%%%

\begin{figure}[h!]
\centering
  \epsfig{file=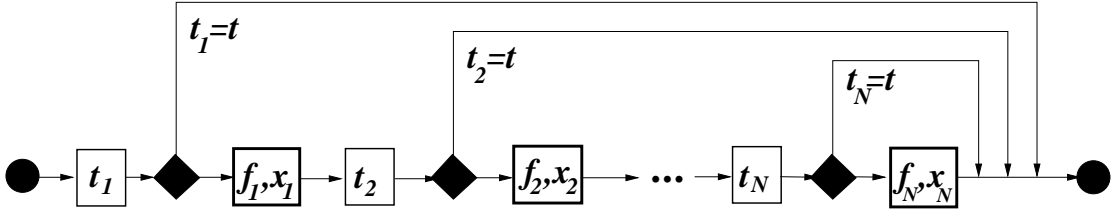,width=150mm}
\caption{\sf
Scheme of the standard Markovian Monte Carlo.
  }
\label{fig:mmc1}
\end{figure}

In the Markovian MC we are going to generate
$t_i$, one after another, starting from $t_1$
until for certain $n$, $0\leq n \leq N$,
$t=t_{n+1}$ is reached%
\footnote{Maximum number of steps $N$ is large and fixed.
   Formally, $N\to\infty$ is understood.}.
For this to be feasible in the Markovian MC, we have to show
with the same algebra as in eq.~(\ref{eq:evol5}),
that all integrals over $t_i$ are properly normalized to
momentum fraction%
\footnote{Unitary normalization is obtained by means
  of applying $1/x_{i-1}$ normalization factor.}
$x_{i-1}$, starting with the innermost $\int_{t_{N-1}}^{t} dt_{N}$
and finishing with outermost $\int_{t_{0}}^{t} dt_1$.
Following the above warm-up example one can show that the integration
over $t_1$ decouples completely from all inner integrations over
$t_2,...,t_N$ and, therefore, can be generated independently
as a first variable in the MC algorithm.

In the MC generation, whenever $\delta_{t_{n+1}=t}$ term
is encountered for the first time,
the real parton emission chain is terminated.
More precisely, for all $k>n$
one may formally define $t_k=t$, but they are dummy (not used).

In ref.~\cite{mcguide:1999} it was stated,
that every standard (classic) MC algorithm
can be reduced to a superposition of only three elementary methods:
mapping of variables, weighting-rejecting and branching.
As seen in Fig.~\ref{fig:mmc1}, where the above basic MMC algorithm
is depicted using graphical notation of ref.~\cite{mcguide:1999},
it is indeed a superposition of branching and mapping --
every box \fbox{$f_i,x_i$} typically includes more elementary methods
(typically mappings and branchings).

What still remains is to define in a more detail
the distribution of all three variables of $t_i,f_i,x_i$ 
of the single Markovian step
after generating $t_{i-1},f_{i-1},x_{i-1}$ in the preceding step:
\begin{equation}
\label{eq:omega}
\begin{split}
1&= \frac{1}{x_{i-1}}
\int\limits_{t_{i-1}}^{t} dt_i\;
\Big\{
  \bar{\bf E}{\bf K}^R(t_{i})
   {\bf G}_{{\bf K}^V}(t_{i},t_{i-1})
  +\bar{\bf E}{\bf G}_{{\bf K}^V}(t,t_{i-1})\delta_{t_{i}=t}
\Big\}_{f_{i-1}}(x_{i-1})
\\&
=\frac{1}{x_{i-1}}
\int\limits_{t_{i-1}}^{t} dt_i
\bigg\{\Big[
 \sum_{f_i}
 \int dx_i\; x_i
  \Keu^R_{f_if_{i-1}}(t_{i},x_i,x_{i-1})
   e^{-\Phi_{f_{i-1}}(t_{i},t_{i-1})}
\Big]
  +x_{i-1}\delta_{t_{i}=t}
   e^{-\Phi_{f_{i-1}}(t,t_{i-1})}
\bigg\}
\\&
=\int\limits_{t_{i-1}}^{t} dt_i
 \sum_{f_i}
 \int dx_i\;
\omega(t_i,f_i,x_i|t_{i-1},f_{i-1},x_{i-1}).
\end{split}
\end{equation}
Let us also show the above
distribution in a form immediately suitable for the MC generation
\begin{equation}
\label{eq:probstep}
\begin{split}
1&=
  e^{-\Phi_{f_{i-1}}(t,t_{i-1})}
+\int\limits^1_{e^{-\Phi_{f_{i-1}}(t,t_{i-1})}}
 d\left(e^{-\Phi_{f_{i-1}}(t_{i},t_{i-1})}\right)
\\& ~~~~\times
 \left[
 \sum_{f_i}
 \frac{\partial_{t_i}\Phi_{f_i f_{i-1}}(t_{i},t_{i-1}|x_{i-1})}%
      {\partial_{t_i}\Phi_{f_{i-1}}(t_{i},t_{i-1}|x_{i-1})}
 \int dx_i\;
 \frac{1}{\partial_{t_i}\Phi_{f_i f_{i-1}}(t_{i},t_{i-1}|x_{i-1})}
 \frac{x_i}{x_{i-1}}
 \Keu^R_{f_if_{i-1}}(t_{i},x_i,x_{i-1})
 \right],
\end{split}
\end{equation}
where virtual form-factor is evaluated using real emission
kernels and split into contributions from various
transition channels according to
\begin{equation}
\label{eq:phireal}
\begin{split}
&\Phi_f (t_1, t_0 |u)
  =\int\limits^{t_1}_{t_0} d t\; {\Keu^v_{ff} (t, u)} 
  =\sum_{f'} 
   \int\limits^{t_1}_{t_0} d t 
   \int\limits_0^u \frac{d x}{u} x\; \Keu^R_{f'f}(t,x,u)
  =\sum_{f'} \Phi_{f'f} (t_1, t_0 |u).
\end{split}
\end{equation}
Given an uniform random number $r\in(0,1)$,
generation of $t_i$ is done by means of solving the equation
$r=U(t_i)=e^{-\Phi_{f_{i-1}}(t_{i},t_{i-1})}$ for $t_i$,
within the range $r\in [e^{-\Phi_{f_{i-1}}(t,t_{i-1})},1]$.
The remaining range $r\in [0,e^{-\Phi_{f_{i-1}}(t,t_{i-1})}]$
is mapped into a single point $t_i=t$, that is the point
where the distribution proportional to $\delta_{t=t_i}$ resides.
Flavour index $f_i$ is generated according 
to normalized discrete probability distribution
$P_{f_i}=\partial_{t_i}\Phi_{f_i f_{i-1}}(t_{i},t_{i-1})/
         \partial_{t_i}\Phi_{f_{i-1}}(t_{i},t_{i-1})$.
Finally, variable $x_i$ is generated
according to the normalized integrand of $\int dx_i$
in eq.~(\ref{eq:probstep}).

The above Markovian MC algorithm of Fig.~\ref{fig:mmc1}
is completely standard and very well known.
Practical problem is that the generation of $t_i$,
for more complicated kernels than in DGLAP case requires
numerical evaluation and inversion
of the form-factor $\Phi_{f_i f_{i-1}}(t_{i},t_{i-1})$.
Generation of $f_i$ is always rather trivial.
On the other hand, generation of $x_i$ can be also nontrivial.
The above problems can be solved, at least partly,
by more sophisticated versions of the Markovian MC,
generally using MC weights, see next section.

\begin{figure}[h!]
\centering
  \epsfig{file=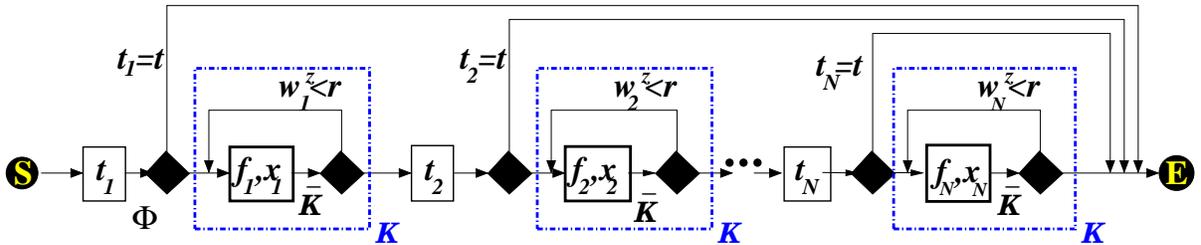,width=160mm}
\caption{\sf
Scheme of Markovian Monte Carlo with the internal rejection loop.
  }
\label{fig:mmc2}
\end{figure}

%%%%%%%%%%%%%%%%%%%%%%%%%%%%%%%%%%%%%%%%%%%%%%%%%%%%%%%%%%%%%%%%%%%%%
\subsection{Weighted Markovian MC algorithms}

In the simplest Markovian MC method with weighted events, 
which will be referred to as an {\em internal loop MMC},
the real emission kernel in the distribution
used in the generation of $x_i$ is replaced by the simplified one
$\Keu^R_{f_if_{i-1}}(t_{i},x_i,x_{i-1})\to
 \bar{\Keu}^R_{f_if_{i-1}}(t_{i},x_i,x_{i-1})$, such that
$\Keu^R \leq \bar{\Keu}^R$.
Variables $x_i$ are generated according to normalized distribution
\begin{equation}
 \bar{P}(x_i)=
 \frac{1}{\partial_{t_i}\bar\Phi_{f_i f_{i-1}}(t_{i},t_{i-1}|x_{i-1})}
 \frac{x_i}{x_{i-1}}
 \bar{\Keu}^R_{f_if_{i-1}}(t_{i},x_i,x_{i-1}),
\end{equation}
where
\begin{equation}
\bar\Phi_{f'f} (t_1, t_0 |u)=
   \int\limits^{t_1}_{t_0} d t 
   \int\limits_0^u \frac{d x}{u} x\; \bar\Keu^R_{f'f}(t,x,u)
\end{equation}
is also simpler than $\Phi$.
The above simplification is corrected by the MC weight
\begin{equation}
w^z_i=\frac{      \Keu^R_{f_if_{i-1}}(t_{i},x_i,x_{i-1})}%
           {\bar{\Keu}^R_{f_if_{i-1}}(t_{i},x_i,x_{i-1})} \leq 1,
\end{equation}
which is used in the local rejection loop,
for every forward step separately, using uniform random number $r$:
if $r>w^z_i$ then generation of $x_i$ is repeated.
In this method generation of $t_i$ is still done using
the exact Sudakov form-factor 
$\Phi_{f_{i-1}}(t_{i},t_{i-1}|x_{i-1})$.
This type of MMC algorithm is shown schematically in Fig.~\ref{fig:mmc2}
and it is essentially a particular realization of the basic
algorithm of Fig.~\ref{fig:mmc1}.
In the second method, which will be referred to as
a {\em global loop MMC} the approximate form-factor
$\bar\Phi_{f_{i-1}}(t_{i},t_{i-1}|x_{i-1})=
 \sum_{f_i} \bar\Phi_{f_i f_{i-1}}(t_{i},t_{i-1}|x_{i-1})$
is used for generation of both $t_i$, $f_i$ and $x_i$.
Global correcting weight $w$ is applied at 
the very end of the Markovian chain.
However, the weight is not just $\prod w^z_i$,
but it can be deduced as follows.
According to eq.~(\ref{eq:probstep}) the normalized probability of the
forward step $(i-1)\to i$ reads
\begin{equation}
\label{eq:omega1}
\begin{split}
&
\frac{dP_{f_i}}{dx_i dt_i}(t_{i-1},f_{i-1},x_{i-1})=
\omega_{(i-1)\to i}=
\omega_{(i-1)\to i}^R+\omega_{(i-1)\to i}^\delta=
\\&
=\theta_{t_{i-1}\leq t_i <t}\;
 \frac{x_i}{x_{i-1}}
  \Keu^R_{f_if_{i-1}}(t_{i},x_i,x_{i-1})
   e^{-\Phi_{f_{i-1}}(t_{i},t_{i-1})}
  +\delta_{t_{i}=t}\delta_{f_if_{i-1}}\delta_{x_i=x_{i-1}}\;
   e^{-\Phi_{f_{i-1}}(t,t_{i-1})}.
\end{split}
\end{equation}
The desired distribution of all variables in MMC event with $n$ emission is
\begin{equation}
\label{eq:omega2}
 \omega^{(n)}=\omega_{n\to n+1}^\delta \prod_{i=1}^n \omega_{(i-1)\to i}^R.
\end{equation}
However, in the actual global loop MMC method the distribution
of these variables (before applying correcting MC weight) is the following
\begin{equation}
 \bar\omega^{(n)}
=\bar\omega_{n\to n+1}^\delta \prod_{i=1}^n \bar\omega_{(i-1)\to i}^R,
\end{equation}
where barring means substitution of exact kernels and form-factors
with the approximate ones:
$\Keu \to \bar\Keu$, $\Phi\to\bar\Phi$.
Global correcting MC weight is, therefore, just the usual ratio
of the exact and approximate distributions
\begin{equation}
w^{(n)}
=\frac{\omega^{(n)}}{\bar\omega^{(n)}}
= e^{\bar\Phi_{f_{n}}(t,t_{n})-\Phi_{f_{n}}(t,t_{n})}
  \left(
  \prod_{i=1}^n w^z_i\; 
  e^{\bar\Phi_{f_{i-1}}(t_{i},t_{i-1})-\Phi_{f_{i-1}}(t_{i},t_{i-1})}
  \right).
\end{equation}

%%%%%%%%%%%%%%%%%%%%%%%%%%%%%%%%%%%%%%%%%%%%%%%%%%%%%%%%%%%%%%%%%%%%%%
\begin{figure}[h!]
\centering
  \epsfig{file=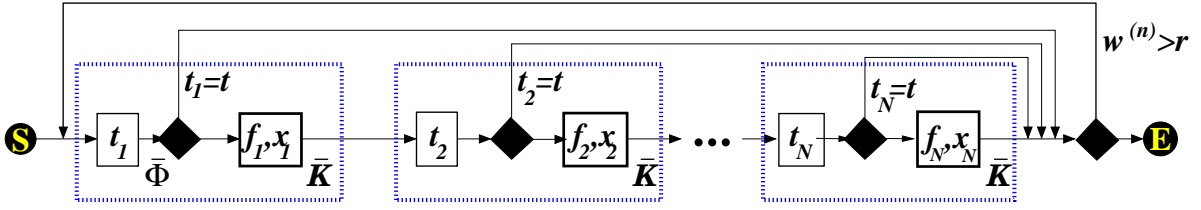,width=160mm}
\caption{\sf
Scheme of Markovian Monte Carlo with the global rejection loop.
  }
\label{fig:mmc3}
\end{figure}
The above weight is tested against the random number after
the entire MC event generation is completed,
see the external return loop in Fig.~\ref{fig:mmc3}.
Note that, although approximate form-factor $\bar\Phi_{f_{i-1}}(t_{i},t_{i-1})$
and its inverse is used here for generation of $t_i$, 
the exact form-factor is still needed to calculate the global weight%
\footnote{ Note that in our older papers describing this method
  we were denoting $T_f=\bar\Phi_f$
  and $\Delta_f=\bar\Phi_f-\Phi_f$.}.

Finally, we are going to derive the third method which will be referred
to as {\em MMC with pseudo-emissions}.
This method is also known in the literature
under the name of the {\em Markovian MC algorithm with veto}
or shortly {\em veto algorithm}.
In this case we do the following modification
of the evolution kernel
\begin{equation}
\begin{split}
&\tilde\Keu^V_{ff'}(t,x,u)=
 \Keu^V_{ff'}(t,x,u)-\delta_{ff'}\delta_{x=u}\Keu^S_{ff}(t,x),
\\&
\tilde\Keu^R_{ff'}(t,x,u)=
\Keu^R_{ff'}(t,x,u)+\delta_{ff'}\delta_{x=u}\Keu^S_{ff}(t,x),
\end{split}
\end{equation}
where $\Keu^S_{ff}(t,x)$ is positive and its magnitude
is judiciously chosen as the integral difference
of the exact kernel $\Keu^R$ and the approximate kernel
$\bar\Keu^R\geq \Keu^R$
(typically the same as in the previous methods)
\begin{equation}
 \Keu^S_{ff}(t,u)=
 \sum_{f'} \int_0^1 dx\; \frac{x}{u}
 \left(\bar\Keu^R_{f'f}(t,x,u)-\Keu^R_{f'f}(t,x,u)\right).
\end{equation}
In this way we are artificially adding to the real emission kernel
finite positive contributions, which represents real emission of
a gluon with exactly zero momentum!
This extra real emission is compensated immediately
and exactly by enlarging negative virtual correction.
Since the total evolution kernel remains unchanged,
\begin{equation}
\Keu_{ff'}(t,x,u)= \tilde\Keu^V_{ff'}(t,x,u)+\tilde\Keu^R_{ff'}(t,x,u),
\end{equation}
the same time-ordered exponential solution remains valid,
${\bf D}(t)={\bf G}_{\bf K}(t,t_0) {\bf D}(t_0)$.
However, the difference will occur when resumming virtual negative
corrections, because we are now resumming the enlarged $\tilde\Keu^V$.
The basic solution used as a starting point for MMC now reads
\begin{equation}
\begin{split}
&
{\bf D}(t)=
\sum_{n=0}^\infty
 \left[\prod_{i=1}^n \int_{t_{i-1}}^{t}dt_i\right]
 {\bf G}_{\tilde{\bf K}^V}(t,t_{n})\; 
 \left[\prod_{i=1}^n
    \tilde{\bf K}^R(t_{i})
    {\bf G}_{\tilde{\bf K}^V}(t_{i},t_{i-1})
 \right]
{\bf D}(t_0),
\\&
\{{\bf G}_{\tilde{\bf K}^V}(t,t')\}_{ff'}(x,u)
  =\delta_{ff'}\delta_{x=u}\; e^{-\tilde\Phi_f(t,t'|x)},\qquad
\tilde\Phi_f(t,t'|x)=\int_{t'}^{t} dt''\; \tilde\Keu^v_{ff}(t'',x).
\end{split}
\end{equation}

The momentum sum rule still holds and can be used to evaluate
modified form-factor
\begin{equation}
\begin{split}
\tilde\Phi_f (t_1, t_0 |u)
& =\int\limits^{t_1}_{t_0} d t\; {\tilde\Keu^v_{ff} (t, u)} 
  =\int\limits^{t_1}_{t_0} d t
   \sum_{f'} 
   \int\limits_0^u \frac{d x}{u} x\; \tilde\Keu^R_{f'f}(t,x,u)
%\\&
%  =\int\limits^{t_1}_{t_0} d t
%   \sum_{f'} 
%   \int\limits_0^u \frac{d x}{u} x\;
%   [\Keu^R_{f'f}(t,x,u)+\delta_{ff'}\delta_{x=u}\Keu^S_{ff}(t,u)]
\\&
  =\int\limits^{t_1}_{t_0} d t
   \left(
   \sum_{f'} 
   \int\limits_0^u \frac{d x}{u} x\;
   \Keu^R_{f'f}(t,x,u)
  +\Keu^S_{ff}(t,u)
   \right)
\\&
  =\int\limits^{t_1}_{t_0} d t
   \sum_{f'} 
   \int\limits_0^u \frac{d x}{u} x\;
   \bar\Keu^R_{f'f}(t,x,u)
  =\bar\Phi_f (t_1, t_0 |u).
\end{split}
\end{equation}
Obviously, $\Keu^S$ was adjusted such that
$\tilde\Phi_f=\bar\Phi_f$ holds.
The immediate important gain is that simplified form-factor $\bar\Phi_f$ 
is used to generate $t_i$, instead of more complicated $\Phi_f$.

%%%%%%%%%%%%%%%%%%%%%%%%%%%%%%%%%%%%%%%%%%%%%%%%%%%%%%%%%%%%%%%%%%%%%
\begin{figure}[h!]
\centering
  \epsfig{file=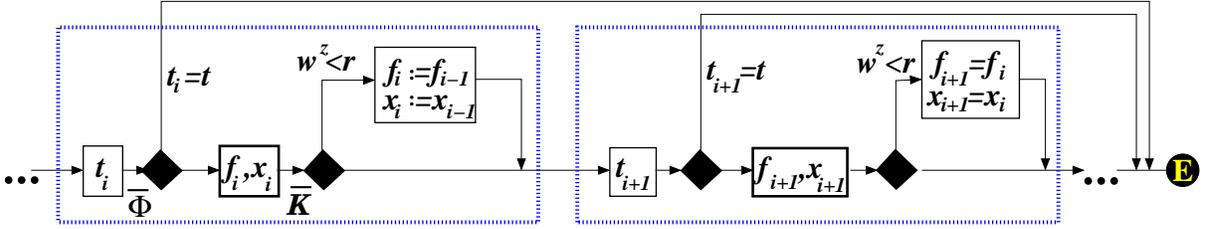,width=160mm}
\caption{\sf
Scheme of Markovian Monte Carlo with pseudo-emissions (veto).
  }
\label{fig:mmc4}
\end{figure}
However, there is one more possible gain from $\tilde\Phi_f=\bar\Phi_f$
in the algorithm of generating $f_i$ and $x_i$.
Due to $\Keu\to\tilde\Keu$,
the probability of choosing $f_i$ should be
\begin{equation}
\tilde{P}_{f_i}
=\frac{\partial_{t_i}\tilde\Phi_{f_if_{i-1}}(t_i, t_{i-1} |x_{i-1})}%
      {\partial_{t_i}\tilde\Phi_{f_{i-1}}   (t_i, t_{i-1} |x_{i-1})}
=\frac{\partial_{t_i}\tilde\Phi_{f_if_{i-1}}(t_i, t_{i-1} |x_{i-1})}%
      {\partial_{t_i}\bar\Phi_{f_{i-1}}     (t_i, t_{i-1} |x_{i-1})}
\end{equation}
% Wiesiek, the above is coorect
The next $x_i$ should be generated according to 
$ \tilde\Keu^R_{f_if_{i-1}}(t_i,x_i,x_{i-1})$,
including singular part proportional
to $\delta_{x_i=x_{i-1}}\delta_{ff'}$.
However, generating $x_i$ and $f_i$ 
according to this distribution can be inconvenient
and the following clever trick may be helpful.
Let us consider for a moment the internal loop MMC algorithm with 
$\bar{P}_{f_i}=\partial_{t_i}\bar\Phi_{f_if_{i-1}}/\partial_{t_i}\bar\Phi_{f_{i-1}}$ 
for which $x_i$ is generated according to $\bar\Keu(x_{i},\dots)$.
Give uniform random number $r$,
the fraction of MC events obeying $r>w^z_i$ 
will be $(\partial_{t_i}\bar\Phi_{f_{i-1}}-\partial_{t_i}\Phi_{f_{i-1}})/
\partial_{t_i}\bar\Phi_{f_{i-1}}$.
Now, due to $\partial_{t_i}\tilde\Phi_f=\partial_{t_i}\bar\Phi_f$
this fraction happens to be exactly the same as the fraction of events 
$(\partial_{t_i}\tilde\Phi_{f_{i-1}}-\partial_{t_i}\Phi_{f_{i-1}})/
\partial_{t_i}\bar\Phi_{f_{i-1}}$
located in the $\delta_{x_i=x_{i-1}}\delta_{ff'}$ term!

One can therefore proceed almost exactly
as in the internal loop MMC algorithm,
that is generate $f_i$ according to $\bar{P}_i$ and $x_i$
according to kernel $\bar\Keu$, and next, for events with $r>w^z_i$,
instead of repeating generation of $f_i$ and $x_i$ for the same $t_i$,
one sets $f_i=f_{i-1}$ and $x_i=x_{i-1}$ (zero momentum real gluon!)
and proceeds to generation of the next $t_{i+1}$.
This completes description and derivation of the algorithm of
MMC with {\em pseudo-emissions}.
The advantage of this algorithm is that the numerical evaluation
and inversion of the possibly complicated exact form-factor
$\Phi_{ff'}(t,t'|u)$ is not required --
only the simplified version $\bar\Phi_{ff'}(t,t'|u)$ is used.
This type of MMC algorithm with pseudo-emissions
is shown schematically in Fig.~\ref{fig:mmc4}.

Comparing to other derivations of the veto MMC, in our derivation
we reduce veto MMC to the standard MMC without the need of
repetition of the the explicit resummation of the contributions
form $\bar{G}$s (which is typically done in the derivations
of veto MMC in the literature).
We believe that the proof presented here is both simpler and more rigorous.

Finally let us comment on one purely technical point.
One may get false impression that the above algorithm with pseudo-emissions
visualized in Fig.~\ref{fig:mmc4}
cannot be reduced to a superposition of the three elementary methods
of ref.~\cite{mcguide:1999}.
In fact it can be done rather easily --
the above algorithm is just
a variant of the basic algorithm of Fig.~\ref{fig:mmc1},
in which the branch with $W^z<r$ representing emission of another
type of real gluon $\bar{G}$ with exactly zero momentum is present.

\section{Kernels and form-factors}
%%%%%%%%%%%%%%%%%%%%%%%%%%%%%%%%%%%%%%%%%%%
Our main interest is in the CCFM-like evolution with the
evolution time being rapidity and running coupling constant
$\alpha_S$ dependent
on the transverse momentum of the emitted gluon.
The LL DGLAP will be shown as a reference case, while another
with rapidity ordering and $z$-dependent $\alpha_S$ will be also discussed.
as a useful intermediate case between CCFM and DGLAP.
Running coupling constant
\begin{equation}
 \alpha_S(q) = 
 \alpha_S^{(0)}(q)=\frac{2\pi}{\beta_0}\; \frac{1}{\ln q-\ln\Lambda_0}
\label{eq:alpha}
\end{equation}
is taken in the LL approximation.
All three types of evolution in this work are essentially the same as in 
ref.~\cite{Jadach:2007singleCMC}, so we shall reduce to a minimum
presentation of the corresponding three kernels and form-factors.

\subsection{Kinematics}
%%%%%%%%%%%%%%%%%%%%%%%%%%%%%%%%%%
\label{sec:kinematics}
As already stressed we define explicit mapping of the evolution
variables to four-momenta, because of possible applications
in the parton shower MCs.
It will be the same as in ref.~\cite{Jadach:2007singleCMC}
and is basically that of CCFM model~\cite{CCFM}.
We define $k_i^\mu$ to be the momenta of
emitted partons, whereas $q_i^\mu$ denote the virtual partons along the
emission tree.
The initial hadron carries $q^+_h = 2E_h$. 
For each emitted parton we define
\begin{align}
k_i^+ = q_{i-1}^+ -q_i^+ =2E_h (x_{i-1}-x_i)=2E_h x_{i-1}(1-z_i);\;\;\;
\eta_i=\frac{1}{2}\ln\frac{k_i^+}{k_i^-}.
\end{align}
Consequently, the transverse momentum of emitted massless parton reads
\begin{align}
\label{eq:tevol}
k^T_{i}=\sqrt{k_i^+k_i^-} = k_i^+ e^{-\eta_i} = x_{i-1}(1-z_i)2E_he^{-\eta_i}.
\end{align}
This suggests the convenient definition of the rapidity-based
evolution time 
%$t=\log q$ 
as
\begin{align}
t_i = 
%\log |q^2_i|=
-\eta_i +\ln(2E_h)\,.
\end{align}
Now, the transverse momentum of the emitted parton (in units of $1~{\rm GeV}$) becomes:
\begin{align}
k^T_{i} = \eto^{t_i}x_i(1-z_i)/z_i = \eto^{t_i}x_{i-1}(1-z_i) = \eto^{t_i}(x_{i-1}-x_i).
\end{align}

\subsection{Three types of kernels}
\label{sec:threekern}
%%%%%%%%%%%%%%%%%%%%%%%%%%%%%%%%%%%%%%%%%%%%%%%
In the following we are going to
define matrix elements of the kernels
\begin{equation}
({\bf K})_{ff'}(x,u)=
\Keu_{ff'}(t,x,u)= \Keu^V_{ff'}(t,x,u)+\Keu^R_{ff'}(t,x,u),
\end{equation}
starting with the real emission part $\Keu^R_{ff'}(t,x,u)$.
It includes implicitly IR cut-off $u-x>\Delta(x,u)$.
The virtual part $\Keu^V_{ff'}(t,x,u)$
will be determined unambiguously by imposing momentum sum rule.
It includes implicitly $\delta_{ff'}\delta_{x=u}$.
We will use as a basic building block
the real emission part of the LL DGLAP kernel.
In order to facilitate numerical calculation
it is decomposed as follows
\begin{equation}
z P^{(0)}_{f'f}(z)=
  \delta_{f'f} \left(\frac{A_{ff}}{1-z} + F_{ff}(z) \right)
 +(1-\delta_{f'f})F_{f'f}(z),
\end{equation}
($z=x/u$), with the coefficients $A_{ff}$ and functions $F_{f'f}(z)$ defined
in ref.~\cite{Golec-Biernat:2006xw}.
Let us start with pure bremsstrahlung case, real emission part.

{\em Case (A)}: DGLAP LL is introduced here as a reference case:
\begin{equation}
\label{eq:kernBremsA}
\Keu^{R(A)}_{ff} (t, x, u)
 =\frac{\alpha_S (Q_0\eto^t)}{\pi}  \frac{1}{u}
    P_{ff}^{(0)}(x/u)\; \theta_{u-x\geq u\epsilon},
\end{equation}
where $\epsilon$ is infinitesimally small and $z=x/u$.

{\em Case (B)}: The argument in $\alpha_S$ is $(1-z)q=(1-z)e^t=k^T/u$;
as advocated in ref.~\cite{Amati:1980ch}.
For the IR cut-off we use $\Delta(t,u)=\lambda ue^{-t}$:
\begin{equation} 
\label{eq:kernBremsB}
\Keu^{R(B)}_{ff} (t, x, u)
 =\frac{\alpha_S^{(0)}((1-x/u)e^t)}{\pi}  \frac{1}{u}
    P_{ff}^{(0)}(x/u)\; \theta_{u-x\geq u\lambda e^{-t}}\; .
\end{equation}

{\em Case (C)}:  
The coupling constant $\alpha_S$ depends on
the transverse momentum $k^T=(u-x)e^t$, while for
an IR cut-off we choose $\Delta(t,u)=\Delta(t)=\lambda e^{-t}$.
The kernel reads:
\begin{equation}
\label{eq:kernBremsC}
\Keu^{R(C)}_{ff} (t, x, u)
 =\frac{\alpha_S^{(0)} ((u-x)e^t)}{\pi}  \frac{1}{u}
    P_{ff}^{(0)}(x/u)\theta_{u-x\geq \lambda e^{-t}}.
\end{equation}

The generalized kernels beyond the case of the pure bremsstrahlung,
for the quark-gluon transitions,
valid for all three cases $X=A,B,C$,
we define as follows
\begin{equation}
\label{eq:kernGen}
x\Keu^{R(X)}_{f'f}(t,x,u)
 = \delta_{f'f}\;
   x\Keu^{R(X)}_{f'f}(t,x,u)
 +(1-\delta_{f'f}) \frac{\alpha_S (e^t)}{\pi} F_{f'f}(z)
     \theta_{u-x>\Delta^{(X)}(u)},
\end{equation}
where $\alpha_S$ in the flavour changing elements have no
$z$- or $k^T$-dependence and the IR cut-off $\Delta^{(X)}$ is the same as
in the bremsstrahlung case.

Note that the case (C) is fully compatible with the CCFM evolution~\cite{CCFM},
except that for the gluon gluon transitions (bremsstrahlung)
the non-Sudakov form-factor assuring the compatibility
with BFKL \cite{BFKL} is not shown
(although it is already present in the MC program)%
\footnote{The original CCFM was formulated for pure gluonstrahlung,
   without quark gluon transitions.}.

As in ref.~\cite{Jadach:2007singleCMC}, for cases (B) and (C),
we also introduce slightly modified version of the
quark-gluon changing kernels elements:
\begin{equation}
\label{eq:kernGen1}
\begin{split}
&x\Keu^{R(B')}_{f'f}(t,x,u)
 = \delta_{f'f}\; x\Keu^{R(B)}_{f'f}(t,x,u)
 +(1-\delta_{f'f}) \frac{\alpha_S ((1-z)e^t)}{\pi} F_{f'f}(z)
  \theta_{1-z>\lambda e^{-t}}\; ,
\\
&x\Keu^{R(C')}_{f'f}(t,x,u)
 = \delta_{f'f}\; x\Keu^{R(C)}_{f'f}(t,x,u)
 +(1-\delta_{f'f}) \frac{\alpha_S (u(1-z)e^t)}{\pi} F_{f'f}(z)
  \theta_{u-x>\lambda e^{-t}},
\end{split}
\end{equation}
with the same arguments of $\alpha_S$ and IR cut-off
as for gluonstrahlung.
New variants are referred to as cases (B') and (C').
One can go back from cases (B') and (C') to (B) and (C)
by means of applying well behaving MC weight.

\subsection{Form-factors}
%%%%%%%%%%%%%%%%%%%%%%%%%%%%%%%%%
Sudakov form-factor resulting from resummation of the virtual
part in the kernel was defined in eq.~(\ref{eq:phidef}).
The virtual part of the kernel is determined through momentum
sum rule, see eq.~(\ref{eq:kernv}), 
leading to the following expression
\begin{equation}
\label{eq:formfactor}
\begin{split}
&\Phi_f (t_1, t_0 |u)
  = \sum_{f'} \int^{t_1}_{t_0} d t 
    \int_0^u \frac{d x}{u} x\; \Keu^R_{f'f}(t,x,u)
\\& 
=\sum_{f'} \int^{t_1}_{t_0} d t 
   \int_0^u \frac{d y}{u} (u-y)\; 
    \Keu^R_{f'f}(t,u-y,u)
%\\& 
  =\sum_{f'} \int^{t_1}_{t_0} d t 
   \int_0^1 dz\; uz \Keu^R_{f'f}(t,uz,u),
\end{split}
\end{equation}
where $z\equiv x/u$ and $y\equiv u-x = (1-z)u$,
see also eq.~(\ref{eq:phireal}).

Following decomposition of the LL kernel into three parts
\begin{equation}
z P^{(0)}_{f'f}(z)=
  \delta_{f'f}\frac{A_{ff}}{1-z} + \delta_{f'f}F_{ff}(z)
 +(1-\delta_{f'f})F_{f'f}(z),
\end{equation}
the Sudakov form-factor for practical reasons is
split into three corresponding parts:
\begin{equation}
 \Phi_f(t_1,t_0|u) 
 =\mathbf{\Phi}_f (t_1, t_0|u)
       + \Phi^b_f (t_1, t_0|u)
       + \Phi^c_f (t_1, t_0|u).
\end{equation}

We show in the following explicit expressions for
the above form-factor components for most complicated case (C),
referring the reader to ref.~\cite{Jadach:2007singleCMC}
for simpler cases (A) and (B):
\begin{equation}
\label{eq:PhiC}
\begin{split}
\mathbf{\Phi}_f (t_1, t_0 |u)
& =\int^{t_1}_{t_0} d t\; \int_0^1 dz\;
  \frac{\alpha_S ((1-z)ue^t)}{\pi}
  \frac{A_{ff}}{1-z}
  \theta_{(1-z)u>\lambda e^{-t}}
\\& =A_{ff} \frac{2}{\beta_0} 
      \varrho_2(\tB_0+\ln u,\tB_1 +\ln u;\tB_\lambda),
\\
\Phi^b_f (t_1, t_0 |u) 
& =\int^{t_1}_{t_0} d t\;
  \int_0^1 dz\;
  \frac{\alpha_S ((1-z)ue^t)}{\pi}
  F_{ff}(z)\;
  \theta_{(1-z)u>\lambda e^{-t}}\; ,
\\
\Phi^c_f (t_1, t_0 |u) 
&=\int^{t_1}_{t_0} d t\;
  \frac{\alpha_S (e^t)}{\pi}
  \sum_{f'\neq f} 
  \int_0^1 dz\;
  F_{f'f}(z)\;
  \theta_{(1-z)u>\lambda e^{-t}},
\end{split}
\end{equation}
where $\tB_i\equiv t_i-\ln\Lambda_0$, $\tB_\lambda\equiv t_\lambda-\ln\Lambda_0$,
$t_\lambda \equiv \ln\lambda$, while function $\varrho_2$ is defined
in Appendix of ref.~\cite{Jadach:2007singleCMC} in terms of log functions.
Two other components $\Phi^b_f$ and $\Phi^c_f$ are evaluated 
numerically for every MC event.
This is feasible, provided one integration is performed analytically
(typically that over $v=\ln(1-z)$) and second integration is done
numerically, see ref.~\cite{Jadach:2007singleCMC} for the details.

%-----------------------------------------------------------------------------
\begin{figure}[!ht]
  \centering
  {\epsfig{file=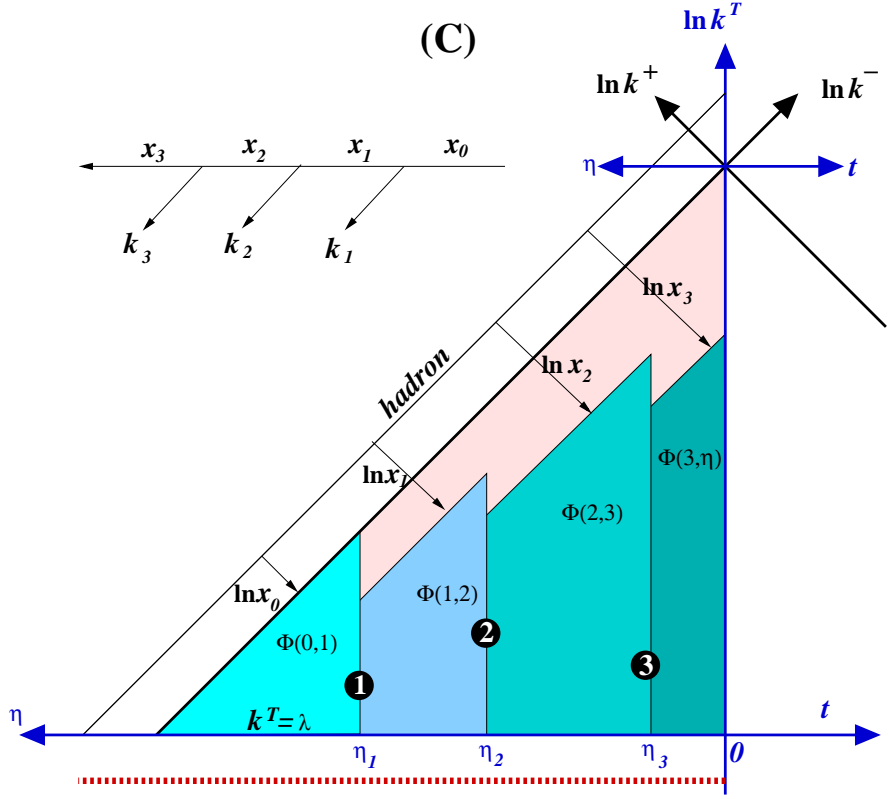, width=120mm}}
  \caption{\sf
  Sudakov plane parametrized two sets of variables
  $(k^+,k^-)$ and $(\eta,\ln k^T)$.
  Emission of three gluons. 
  Their momenta $k^\mu_i, i=1,2,3$ are marked as black numbered circles.
  Position of the Landau pole marked as dashed red line at $k^T=\Lambda_0$.
  Phase space limits as in case (C), that is CCFM evolution.
}%caption
  \label{fig:sudakC}
\end{figure}
%-----------------------------------------------------------------------------

\subsection{Discussion}
%%%%%%%%%%%%%%%%%%%%%%%%%%%%%%%%%%%%%%%%%%%%%%%%%%%
In all three cased (A--C) the distributions of the single forward step
(parton emission) are relatively simple -- they are build out of LL
DGLAP kernels and $\alpha_S$ depending on $t_i$ $z_i$ or $k_T$.
The same distributions enter into form-factor of eq.~(\ref{eq:formfactor}).
Practical problems in the MC implementations are not so much in the
distribution shapes as in the kinematic limits.
We shall therefore concentrate in the following on this subject.
For this purpose we will draw the limits of the available phase space
in the emission of several gluons in the two-dimensional
Sudakov logarithmic plane parametrized with
variables $(k^+,k^-)$ and $(\eta,\ln k^T)$ simultaneously.
The same integration limits are used in the calculation of
the form-factors.
The translation from evolution times
and lightcone variables, $t_i,x_i$, to
rapidities and transverse momenta, $(\eta_i,\ln k^T_i)$,
will be done using mapping of Section.~(\ref{sec:kinematics})
in all three cases (A--C)%
\footnote{This mapping is primarily adequate for (C).
   In principle it could be different for (A) and (B).}.

In in Fig.~\ref{fig:sudakC} we start with case (C).
The total emission phase space has triangular shape
and is limited by maximum rapidity (from right) minimum $k^T$
(from below) and conservation of lightcone plus variable,
$k^+_i<2E_h x_{i-1}$.
Within the above phase space, momenta of three emitted gluons
$k^\mu_i,i=1,2,3$ are represented by the black numbered circles.
They are ordered in rapidity.
The integration domains for the four consecutive
form-factors $\Phi_{f_i}(t_i|t_{i-1})$
in the forward step distributions
in eqs.~(\ref{eq:omega1}--\ref{eq:omega2})
are also shown in Fig.~\ref{fig:sudakC} as a triangle and three trapezoids.

%-----------------------------------------------------------------------------
\begin{figure}[!ht]
  \centering
  {\epsfig{file=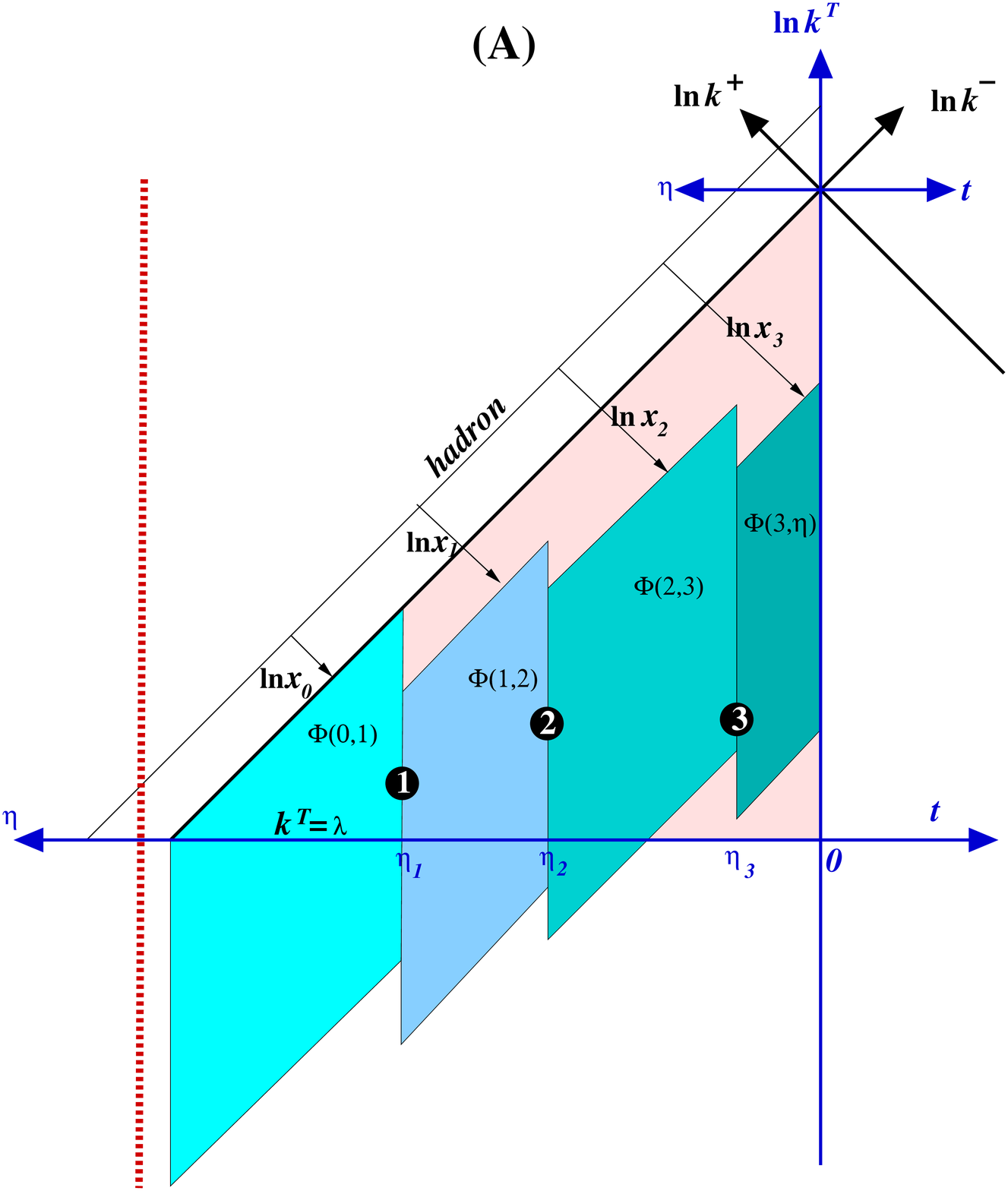, width=75mm}}
  {\epsfig{file=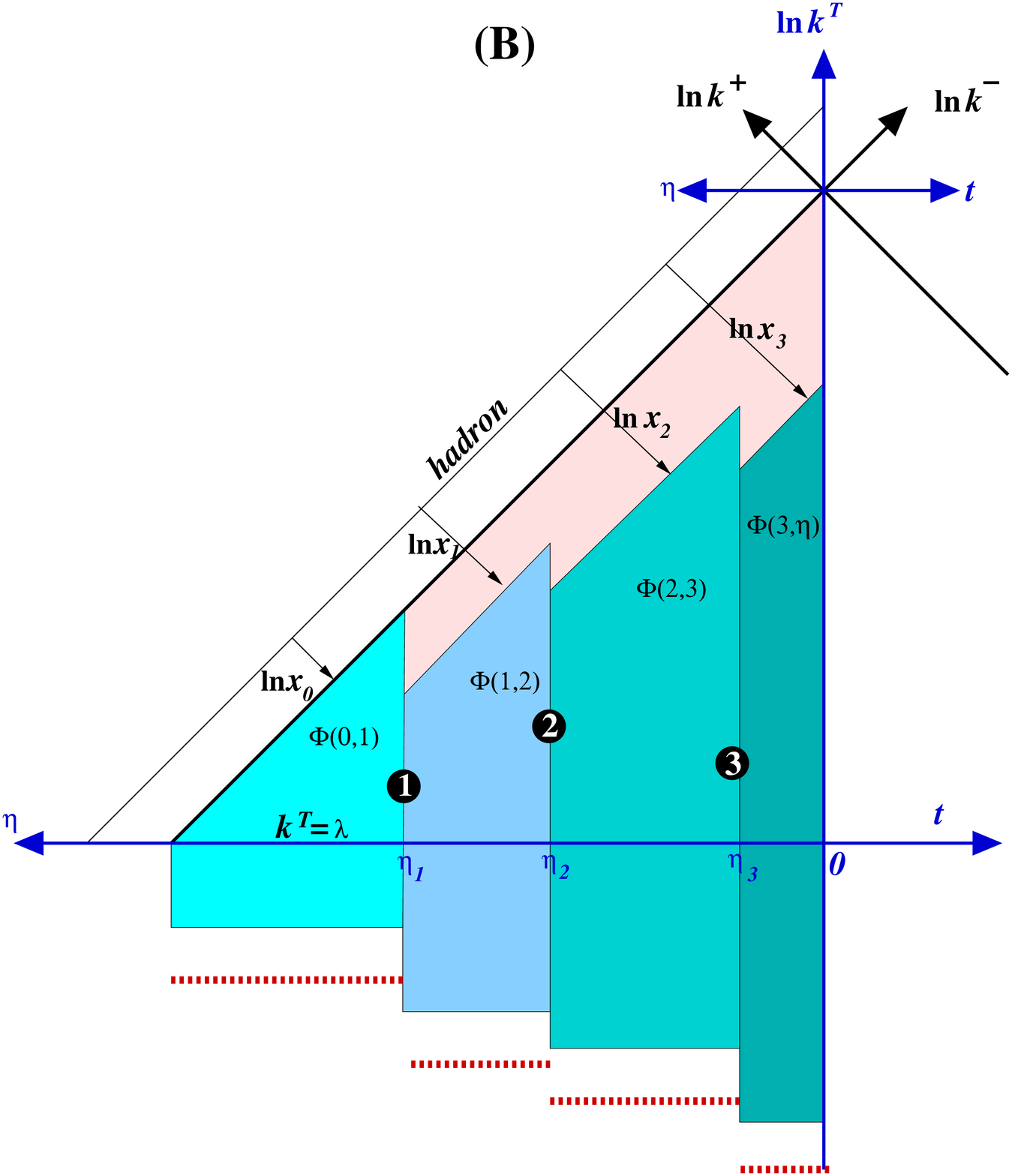, width=75mm}}
  \caption{\sf
  Sudakov plane parametrized with
  $(k^+,k^-)$ and $(\eta,\ln k^T)$.
  Emission of three gluons. 
  Their momenta $k^\mu_i, i=1,2,3$ are marked as black numbered circles.
  Position of the Landau pole marked as dashed red line at 
  (A) $Q=\Lambda_0$ or (B) $k^T=\Lambda_0$.
  Phase space limits as in case (A) and (B)
}%caption
  \label{fig:sudakAB}
\end{figure}
%-----------------------------------------------------------------------------

It is now interesting to compare 
the phase-space limits in the Sudakov
plane between the case (C) and the two other cases (A) and (B).
The corresponding plots are shown in Fig.~\ref{fig:sudakAB}.
The main difference is in the shape of the lower infrared (IR) boundary
of the emission phase space. In the case (A) of DGLAP it is at the same
distance $\ln(1/\veps)$ from the upper limit, hence rhomboid shapes
with the variable widths and constant heights.
In case (B) the IR limit in $k^T$ is lowered by the factor $x_{i-1}$
which grows after every emission, hence we see the trapezoids with the lower
boundary descending deeper and deeper into smaller $k^T$.
The above illustrates also why
the construction of the MMC programs evolution type (B)
served the role of an intermediate step on the way from DGLAP to CCFM.

%-----------------------------------------------------------------------------
\begin{figure}[!ht]
  \centering
  {\epsfig{file=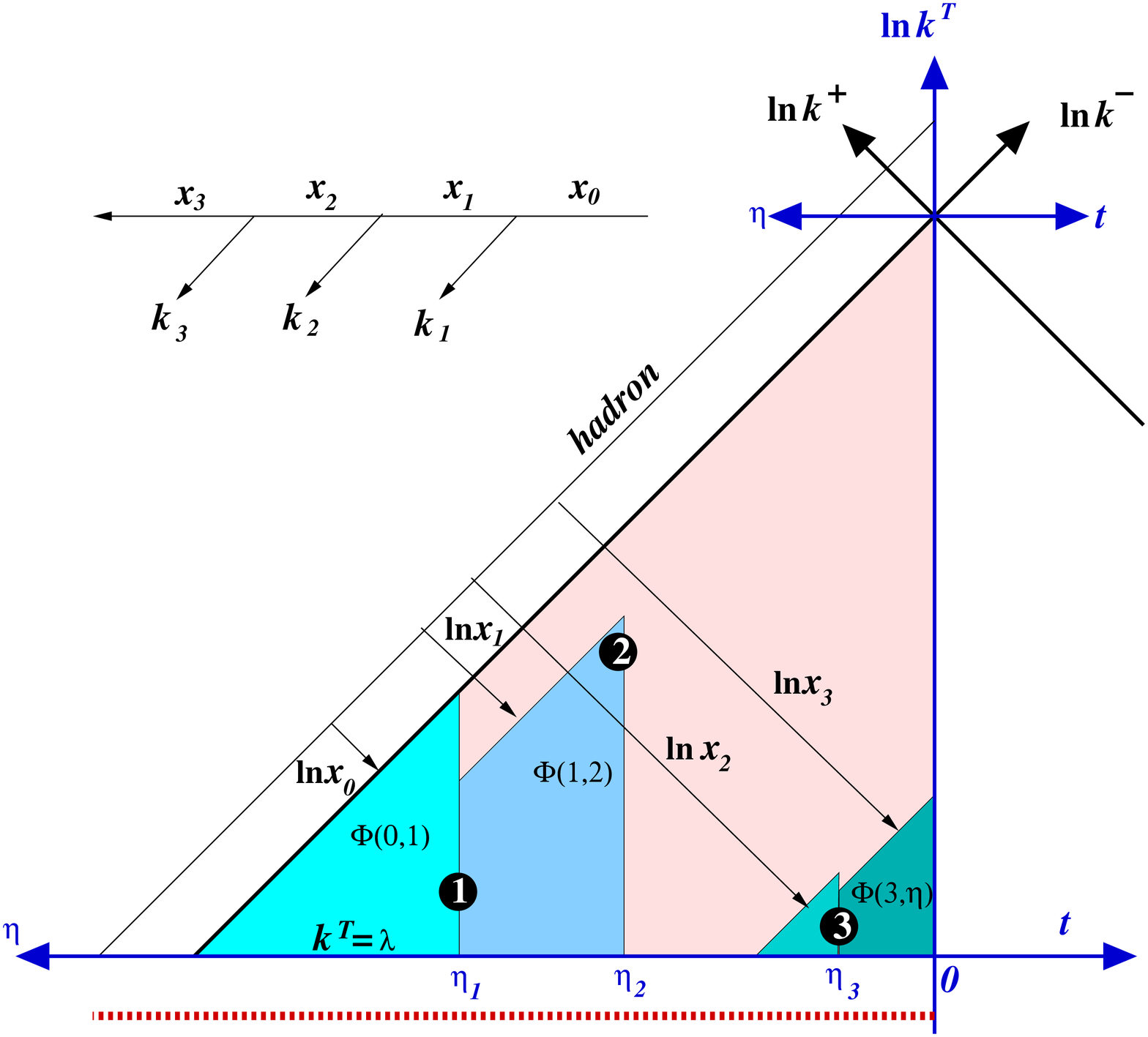, width=120mm}}
  \caption{\sf
  Sudakov plane with emission of three gluons, case (C).
  For $z_2\sim 0$ the second gluon has large $k^T$.
}%caption
  \label{fig:sudakCh}
\end{figure}
%-----------------------------------------------------------------------------
Last not least, let us show kinematic limits in the extreme case of one
$z_i\to 0$. 
This limit is treated in CCFM evolution
better than in DGLAP, because CCFM in this limit
coincides with the BFKL evolution~\cite{BFKL}.
Such a case is illustrated in Fig.~\ref{fig:sudakCh}, where the second emitted
gluon is very hard, that is with high $k^T$, In fact larger than the scale of the
hard process.
(In this part of the phase space the non-Sudakov form-factor
plays significant role.)
The above kinematic region is properly included
in the MMC case (C) and also in the CMC of ref.~\cite{Jadach:2007singleCMC}.

\section{Monte Carlo implementations}
%%%%%%%%%%%%%%%%%%%%%%%%%%%%%%%%%%%%%%%%%%%%%%%%%%%%%%%%%%%%
Studies of the DGLAP evolution, case (A), using the Markovian MCs
were already covered in 
refs.~\cite{Jadach:2003bu,Jadach:2006ku,Placzek:2007xb}
in particular NLO case was extensively studied in
ref.~\cite{Golec-Biernat:2006xw}.
The main aim of these papers was to show that MC method,
although slower, is equally precise and more versatile as compared
to older non-MC techniques,
for example grid method based {\tt QCDnum16}~\cite{qcdnum16}.
These MMCs were also used to test first examples
of the constrained MCs~\cite{Jadach:2005bf,Jadach:2005yq}
for DGLAP-type evolution.
The main advantage of MC method turns out to be
very good and stable estimator of the error.
The slowness of MMCs is mainly the problem in any attempt of
fitting deep-inelastic $ep$ data.
Here, special pretabulation procedures are necessary,
%see refs.~\cite{Jung:2000hk,stoklosa}.
see refs.~\cite{Jung:2000hk}.
The above studies of the evolution type (A)
using MMCs were fairly complete, hence
there is no need to repeat them here.

As already said, we do not show/repeat in this work
tests of MMC type (A)
and we will limit numerical results to comparisons of MMC versus
non-MC program {\tt APCheb} \cite{APCheb40} for evolutions class (B) and (C).
It should be stressed that {\tt APCheb} was originally working only
for DGLAP and was upgraded to evolutions type (B) and (C)
for the purpose of the tests with MMCs.
Comparisons of MMC and CMC programs for evolutions type (B) and (C)
were also done and have been presented in ref.~\cite{Jadach:2007singleCMC}.
In this way we have in our disposal three completely
different programs (sometimes even four) which solve numerically
evolution equations of all three types (A), (B) and (C) and
provide identical results within precision of $0.2\%$!

\subsection{Reusing MMC type (B) as type (C)}
%%%%%%%%%%%%%%%%%%%%%%%%%%%%%%%%%%%%%%%%%%%%%%%
Historically, the MMC for evolution type (B) 
with $\alpha_S(e^t (1 - z))$ and IR cutoff $1-z >\lambda e^{- t}$
was developed first, before CCFM-like scenario (C).
While testing first versions of MMC type (C) the following
observation was helpful.
Examining carefully the propability distributions of the single
forward step $\omega_{(i-1)\to i}$ of eqs.~(\ref{eq:omega1},\ref{eq:probstep})
one may notice that the whole 
additional dependence on the $x_{i-1}$ variable in $\omega^{(C)}_{(i-1)\to i}$ 
can be absorbed into $\lambda$ and $\Lambda_0$:
\begin{equation}
  \omega^{(C)}_{(i-1)\to i}(\lambda,\Lambda_0)=
  \omega^{(B)}_{(i-1)\to i}(\lambda/x_{i-1},\Lambda_0/x_{i-1}).
\end{equation}
Of course, this is the consequence of
the relations $\alpha_S (k_i^T) = \alpha_S (e^{t_i} (1 - z_i) x_{i - 1})$
and $k_i^T = e^{t_i} (1 - z_i) x_{i - 1} > \lambda$.
As a results, we could in the tests of MMC class (C)
{\em reuse} the MMC for \ $\alpha_S (e^t (1 - z))$ by means of reseting 
$\lambda \rightarrow \lambda / x_{i - 1}$ and 
$\Lambda \rightarrow \Lambda / x_{i -1}$,
before generating each single forward step.
The above trick was quite helpful in testing MMC class (C),
for pure bremsstrahlung.

\section{Solving evolution equations with Chebyshev polynomials}
%%%%%%%%%%%%%%%%%%%%%%%%%%%%%%%%%%%%%%%%%%%%%%%%%%%%%%%%%%%%%%%%
In the previous
section the Monte Carlo method for solving the evolution equations was presented. For the sake
of the comparison, we are going to present an alternative method based on the expansion in  the Chebyshev polynomials.

We start from the general form (\ref{eq:genevoleq}) of the evolution equations
\be
\label{eq:num1}
\partial_t D_f(t,x)=\sum_{\fprime}\int_0^1 du\,\Kcal_{f\fprime}(t,x,u)\,D_{\fprime}(t,u)
\ee
with the kernel (\ref{eq:evkernel}).
The momentum sum rule (\ref{eq:momsumrule}) imposed on the parton distributions allows to determine the
virtual part of the kernel  (\ref{eq:evkernel}) from the condition (\ref{eq:kernelsum}).
As a result, we arrive at 
the most general form of the evolution equations
\beeq\nonumber
\partial_t (xD_f(t,x))&=&\sum_{\fprime}\int_0^1 {du}\, x\Kcal_{f\fprime}^R(t,x,u)\,D_{\fprime}(t,u)
\\\label{eq:num7}
&-&D_f(t,x)\,\sum_{\fprime}\int_0^1 du\, u\,\Kcal_{\fprime f}^R(t,u,x)\,.
\eeeq

As an illustration, we consider in detail the evolution equations for the case (C) from Section \ref{sec:threekern}.  
The evolution parameter $t$ in this case
is related to the rapidity $y=\eta$ of the emitted real parton by the relation (\ref{eq:tevol}), which now reads
%(in units of $1~{\rm GeV}$)
\be
k_T=2E_h\eto^{-y}(u-x)=\eto^{t}(u-x)\,.
\ee
Here $u$ and $x$ are the longitudinal momentum fraction before and after the emission.
In the leading logarithmic approximation the real emission kernel takes the form
\be\label{eq:num10}
x\Kcal_{f\fprime}^R(t,x,u)=\frac{\alpha_s(k_T)}{\pi}\,\frac{x}{u}\,
P_{f\fprime}^{\,(0)}\!\!\left(\frac{x}{u}\right)
\theta (u-x)\,,
\ee
where $P_{f\fprime}^{\,(0)}$ are the leading order splitting functions. In order to avoid the Landau pole in $\alpha_s$, we assume that the transverse momenta
of the emitted partons are bounded from below,
\be\label{eq:num11}
k_T\ge \lambda\gg \Lambda_{QCD}\,.
\ee 
This further restricts the momentum fractions $u$ for the real emission (see the theta function in eq.~(\ref{eq:kernBremsC})):
\be\label{eq:num12}
u\ge x+\lambda\eto^{-t}\,.
\ee
Changing the integration variable, $z=x/u$, we obtain for the real emission part of the evolution equations
(\ref{eq:num7})
\be\label{eq:num13}
\sum_{\fprime}\int_x^{z^R(x,t)}dz\,\frac{\alpha_s(\eto^tx(1-z)/z)}{\pi}
\,P_{f\fprime}^{\,(0)}(z)\,\frac{x}{z}D_{\fprime}\!\!\left(t,\frac{x}{z}\right)\,,
\ee
with the upper integration limit given by
\be\label{eq:num14}
z^R(x,t)=\frac{1}{1+\lambda\eto^{-t}}\, >\, x\,.
\ee

For the virtual part of the evolution equations we interchange $u\leftrightarrow x$ in 
the kernel (\ref{eq:num10}). Now, the conditions which restrict the $u$-values read
\be\label{eq:num15}
u<x\,,~~~~~~~~~~~~~k_T=\eto^{t}(x-u)\ge \lambda\,.
\ee
Changing the integration variable, $z=u/x$, in the second integral of eqs.~(\ref{eq:num7}), we obtain
for the virtual term
\be\label{eq:num16}
-xD_f(t,x)\,\sum_{\fprime}\int_0^{z^V(x,t)} dz\, 
\frac{\alpha_s(\eto^tx(1-z))}{\pi}\,zP_{\fprime f}^{\,(0)}(z)\,,
\ee
where now
\be\label{eq:num17}
z^V(x,t)=1-\lambda\eto^{-t}\,>\,0\,.
\ee
In summary, we find the following evolution equations
\beeq\nonumber
\partial_t (xD_f(t,x))\!\!\!&=&\!\!\!
\sum_{\fprime}\int\limits_x^{z^R(x,t)}dz\,\frac{\alpha_s(\eto^tx(1-z)/z)}{\pi}
\,P_{f\fprime}^{\,(0)}(z)\,\frac{x}{z}D_{\fprime}\!\!\left(t,\frac{x}{z}\right)
\\\label{eq:num18}
&-&\!\!\!xD_f(t,x)\,\sum_{\fprime}\int\limits_0^{z^V(x,t)} dz
\frac{\alpha_s(\eto^tx(1-z))}{\pi}\,zP_{\fprime f}^{\,(0)}(z)\,.
\eeeq
These equations are complicated enough to be solved only numerically. In the next section we will
present the method based on the expansion in   the Chebyshev polynomials.

\subsection{Chebyshev polynomial method}

In this method we use the Chebyshev polynomials defined by
\be\label{eq:cheb1}
T_k(y)=\cos(k\arccos(y))\,,~~~~~~~~~~y\in[-1,1]\,.
\ee
The index $k=0,1,2,\ldots$ denotes the polynomial order. Fixing the order, $k=N$, we consider
the equation $T_N(y)=0$. It has $N$ roots (nodes) given by
\be\label{eq:cheb2}
y_i=\cos\frac{\pi}{N}(i-1/2)\,,~~~~~~~i=1,2,\ldots,N\,.
\ee
These roots allow to define the following discrete orthogonality relation for the set of the Chebyshev
polynomials $\{T_0,T_1,\ldots\,T_{N-1}\}$:
\be\label{eq:cheb3}
\sum_{i=1}^N\,T_j(y_i)\, T_k(y_i)=C_j\delta_{jk}\,,
\ee
where $j,k=0,1,\ldots,(N-1)$. The coefficients $C_0=N$ and $C_{j\ge 1}=N/2$. 

A function $f(x)$ with $x\in [a,b]$ can be approximated with the help of the specified set of Chebyshev polynomials in following way
\be\label{eq:cheb4}
f(x)\,\approx\,\sum_{n=1}^N v_n\,c_n\,T_{n-1}(y(x))\,,
\ee
where $v_1=1/2$, $v_{n\ge 1}=1$ and $y=y(x)$  is 
an arbitrary, invertible function which  transforms $[a,b]\to [-1,1]$. The coefficients $c_n$ of the expansion
can be calculated from the orthogonality relation (\ref{eq:cheb3}),
\be\label{eq:cheb5}
c_n=\frac{2}{N}\sum_{i=1}^N f(x_i)\,T_{n-1}(y_i)\,,
\ee
where $x_i=y^{-1}(y_i)$ are images of the roots (\ref{eq:cheb2}) in the interval $[a,b]$.
From relations (\ref{eq:cheb4}) and (\ref{eq:cheb5}) we see that 
one only needs the values $f(x_i)$ at the Chebyshev nodes  to reconstruct
the function at any other $x\in[a,b]$.
This observation is a starting point of the method of the solution of the evolution equations (\ref{eq:num18}).
We simply solve them at the Chebyshev nodes $x=x_k$. 

Therefore, writing eqs.~(\ref{eq:num18}) in a prototype form,
\be\label{eq:cheb6}
\frac{dD(t,x_k)}{dt}=\int_{x_k}^{z(x_k,t)}dz\,P(t,z)\,D(t,x_k/z)\,,
\ee
we consider the finite set of the first order differential equations
for $k=1,2,\ldots,N$. The integration on the r.h.s. needs the values of $D$ at any point, thus we use
the Chebyshev approximation 
\be\label{eq:cheb7}
D(t,x_k/z)\approx \frac{2}{N}\sum_{n=1}^N\sum_{i=1}^N v_n\,D(t,x_i)\,T_{n-1}(y_i)\,T_{n-1}(y(x_k/z))\,.
\ee
Substituting into (\ref{eq:cheb6}), we find the following set of equations
\be\label{eq:cheb8}
\frac{dD(t,x_k)}{dt}=\sum_{i=1}^N\,{\cal{A}}_{ki}(t)\,D(t,x_i)
\ee
which can easily be  solved numerically \cite{APCheb40}. The matrix ${\cal{A}}_{ki}(t)$ in these equations, 
\be\label{eq:cheb9}
{\cal{A}}_{ki}(t)=\frac{2}{N}\sum_{n=1}^N v_n\,T_{n-1}(y_i)\int_{x_k}^{z(x_k,t)}dz\,P(t,z)\,T_{n-1}(y(x_k/z))\,,
\ee
is computed numerically in the process of finding the solution of eqs.~(\ref{eq:cheb8}). 

The differential equations which we consider need initial conditions at some initial scale $D(t=t_0,x)$.
They
are usually specified analytically such that the initial values $D(t_0,x_k)$ at the Chebyshev nodes are easily calculated.

The results of the comparison of the solutions of the evolution equations obtained using the Monte Carlo
and Chebyshev methods are discussed in the next sextion. In general, a very good agreement between the results of these two methods is found.

%-----------------------------------------------------------------------------
\begin{figure}[!ht]
  \centering
  {\epsfig{file=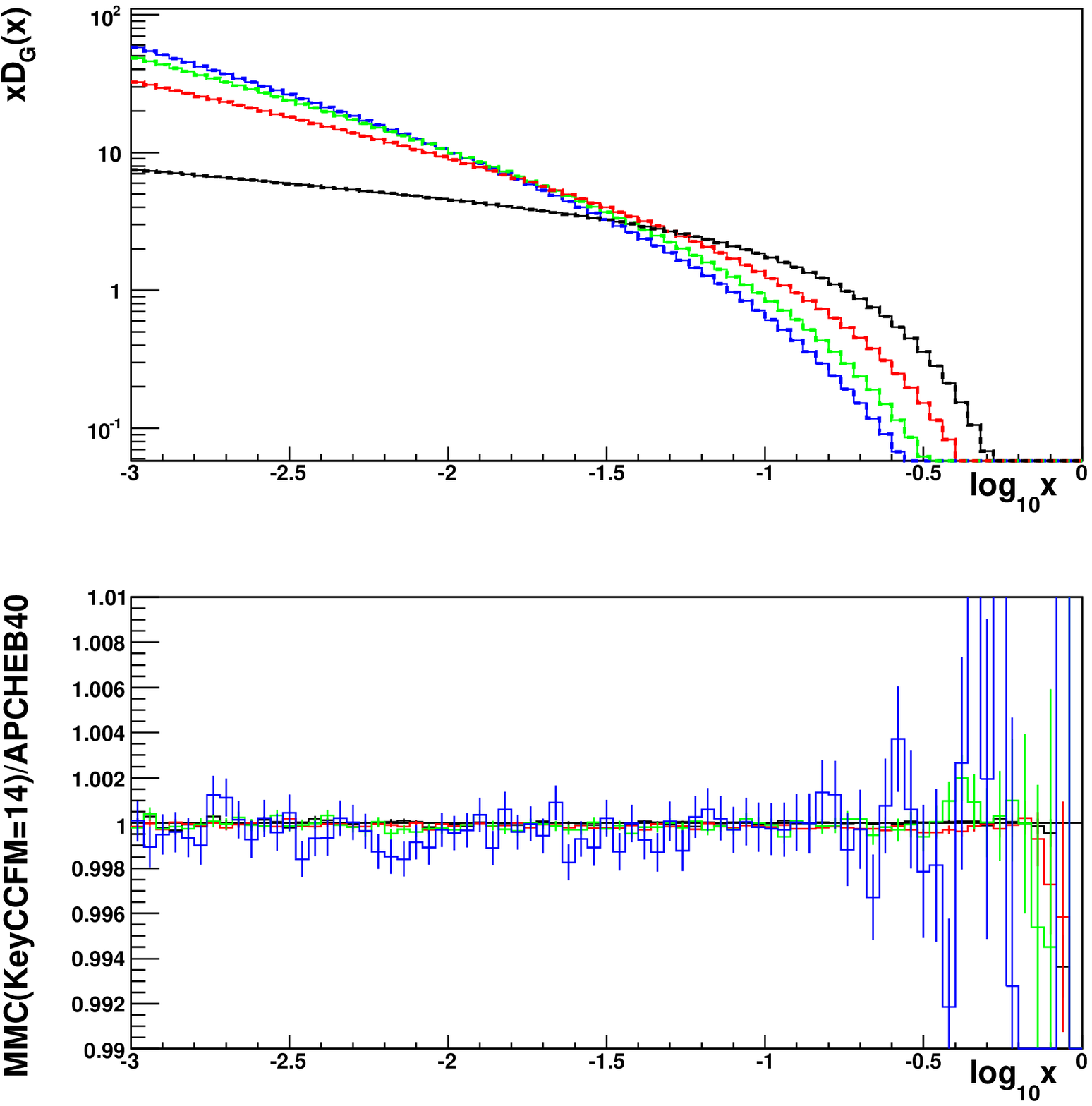, width=100mm}}
  {\epsfig{file=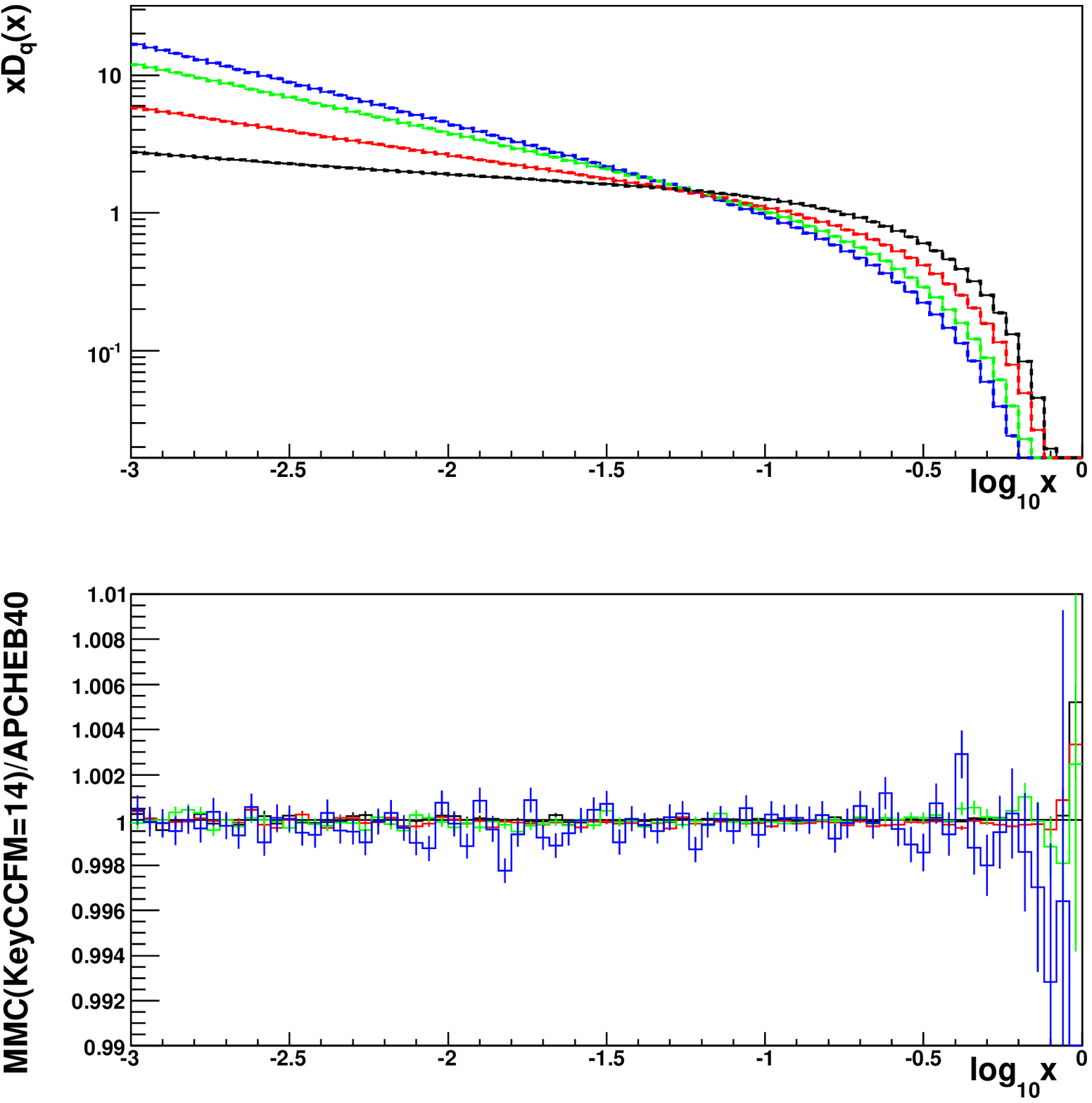, width=100mm}}
  \caption{\sf
    For evolution type (B) with $\alpha((1-z)Q)$
    plotted are distributions $xD_f(x)$ and their ratios
    MMC/APCheb for $f$=gluon (upper) and $f=q+\bar{q}$ (lower plot).
    }
  \label{fig:MMC-APCheb-z}
\end{figure}
%-----------------------------------------------------------------------------

%-----------------------------------------------------------------------------
\begin{figure}[!ht]
  \centering
  {\epsfig{file=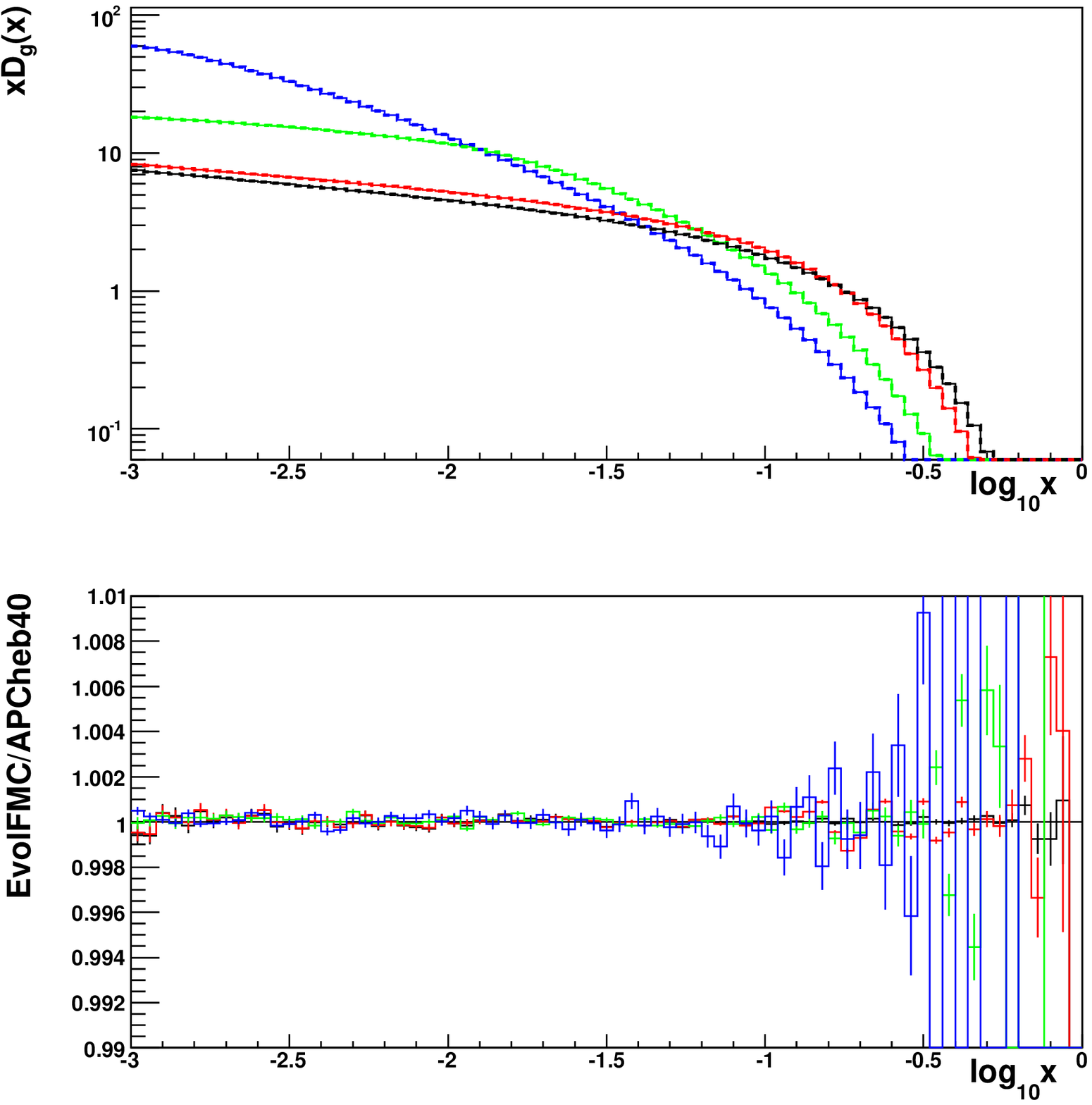, width=100mm}}
  {\epsfig{file=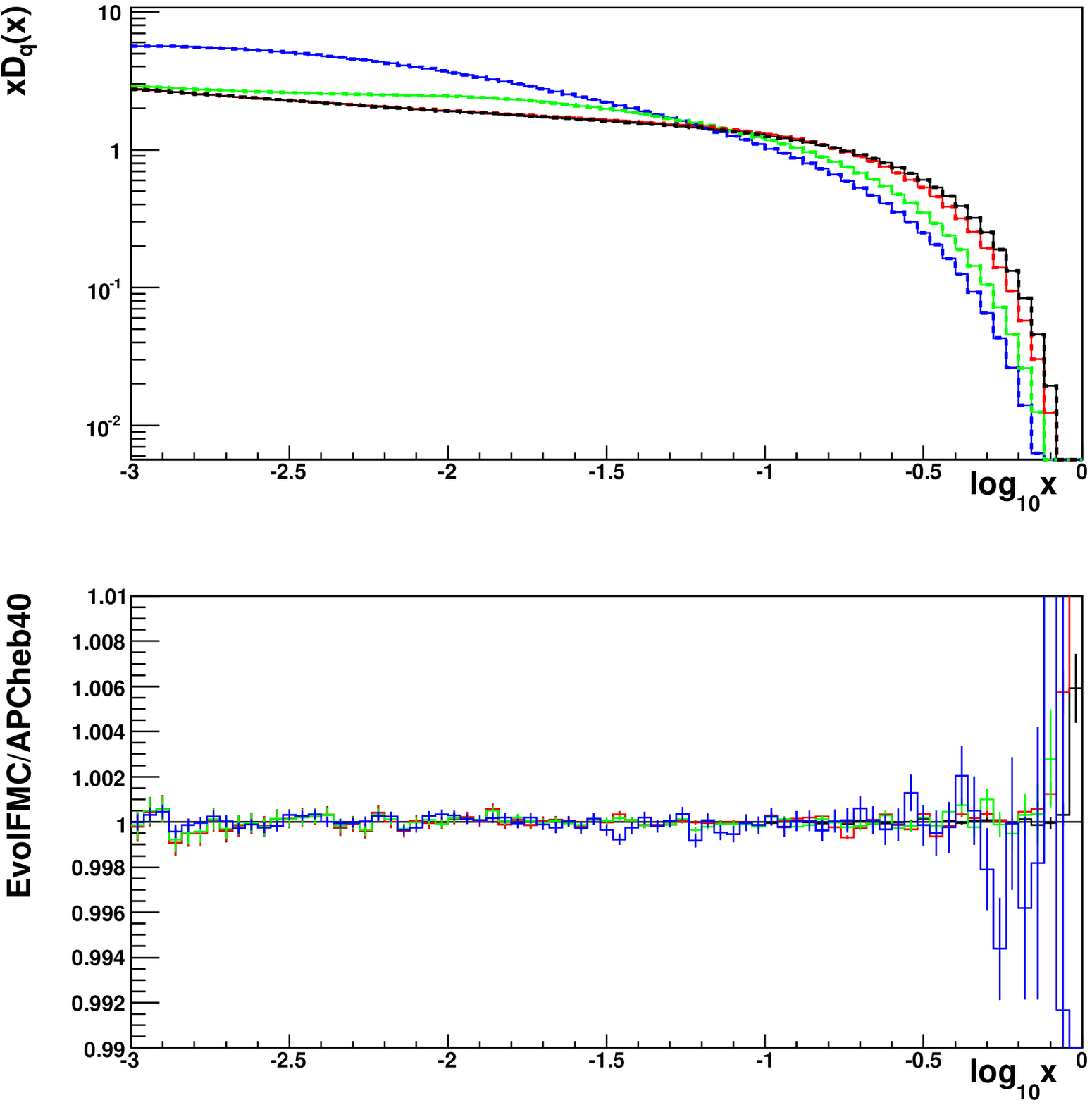, width=100mm}}
  \caption{\sf
    For evolution type (C) with $\alpha(k^T)$
    plotted are distributions $xD_f(x)$ and their ratios
    MMC/APCheb for $f$=gluon (upper) and $f=q+\bar{q}$ (lower plot).
    }
  \label{fig:MMC-APCheb-kT}
\end{figure}

\section{Numerical results}
%%%%%%%%%%%%%%%%%%%%%%%%%%%%%%%%%%%%%%%%%%%%%%%
Although our {\tt MMC} program was 
systematically tested against non-MC programs {\tt APCheb}
and {\tt QCDnum16} for all evolution types (A--C), we shall show
examples of the numerical results for
the more sophisticated and difficult evolution types (B) and (C).

Fig.~\ref{fig:MMC-APCheb-z} demonstrates
distributions $xD_f(t,x)$ from {\tt MMC} and {\tt APCheb}~\cite{APCheb40}
programs and the corresponding ratios {\tt MMC/APCheb}
for the evolution type (B), that is with $\alpha((1-z)Q)$.
The four curves represent $xD_f(t,x)$ for $Q=e^t=1,10,10^2,10^3$GeV.
The upper plots are for $f=G$, gluon while lower plots
are for $f=q+\bar{q}$, quarks and antiquarks taken together.
The starting quark and gluon
distribution at $Q=e^t=1$GeV are defined exactly the
same as in previous works
of refs.~\cite{Jadach:2003bu,Jadach:2006ku,Placzek:2007xb}.
Results for all $Q=1,10,10^2,10^3$GeV were
obtained in the single MC run of $\sim 10^{10}$ MC events.
As we see the distributions from two programs agree within
the statistical MC error of about $~\sim 0.2\%$.

In fig.~\ref{fig:MMC-APCheb-kT} we show the same type of comparison
of {\tt MMC} and {\tt APCheb}, but for evolution type (C).
Again precision agreement within the statistical MC error is reached.

For the LL DGLAP, case  (A), we have reproduced results of 
ref.~\cite{Jadach:2003bu} with smaller statistical errors
and removing certain numerical biases
which were seen in this paper in the gluon case, $f=G$.
We do not show explicitly the corresponding numerical results.

%%%%%%%%%%%%%%%%%%%%%%%%%%%%%%%%%%%%%%%%%%%%%%%%%%%%%%%%%%%%%%%%%%
\section{Summary}
We have developed and tested Markovian MC programs
for two additional types of the QCD evolution equations, in addition to DGLAP.
One of them is identical with the
so-called all-loop CCFM (modulo non-Sudakov form-factor).
The corresponding MC programs were tested to a high-precision level
by means of comparison with the other non-MC program {\tt APCheb}.
MMC of this work is also used to test another class
of the constrained MCs in other independent works,
for the same class to QCD evolutions.
The aim of these exercises is to build basis for the new parton
shower implementations.
The mapping of the evolution
variables into four-momenta was also introduced and tested.

%%%%%%%%%%%%%%%%%%%%%%%%%%%%%%%%%%%%%%%%%%%%%%%%%%%%%%%%%%%%%%%%%%%%%%%%%%%%%
\vspace{10mm}
\noindent
{\bf\Large Acknowledgments}
\vspace{2mm}\\
We would like to thank A. Si\'odmok for useful discussions.
We acknowledge the warm hospitality of the CERN Physics Department, where part
of this work was done. 

%%%%%%%%%%%%%%%%%%%%%%%%%%%%%%%%%%%%%%%%%%%%%%%%%%%%%%%%%%%%%%%%%%%%%%%%%%%%
%\bibliographystyle{utphys_spires}
%\bibliographystyle{h-physrev3}
%\bibliography{radcor}
%%%%%%%%%%%%%%%%%%%%%%%%%%%%%%%%%%%%%%%%%%%%%%%%%%%%%%%%%%%%%%%%%%%%%%%%%%%

\end{document}